# Syn2Logic

## End-to-End Neuromorphic Design Automation

Artur Podobas

Department of Computing and Learning Systems, KTH Royal Institute of Technology, Brinellvägen 8, 100 44 Stockholm, Sweden.

Contributing authors: podobas@kth.se;

**Abstract**

The termination of Dennard's scaling and the end of Moore's law are forcing researchers and computer specialists to consider systems other than classical computers. Among these emerging post-Moore systems are neuromorphic architectures— systems built to resemble and replicate the fantastic computing power of the biological brain. Neuromorphic systems are emerging as a strong candidate to solve problems faster or more energy-efficiently within fields such as computational neuroscience, optimization theory, or artificial intelligence/machine learning. Prior work has demonstrated prototypes that outperform classical systems by nearly three orders of magnitude on certain problems, and much of the success of neuromorphic engineering builds on advances within computational neuroscience.

In this work, we propose a view on electronic Neuromorphic Design Automation (**eNDA**), which we see as a design automation flow that bridges computational neuroscience modeling with traditional Electronic Design Automation (EDA) flow. We introduce the term, give examples of how it can be implemented, and design a prototype implementation: Syn2Logic. Syn2Logic is an entire **eNDA** framework, from language-to-RTL, that allows domain experts to model neural behavior using a custom Domain-Specific Language (DSL) and a compiler that takes the same model description down to synthesizable Register-Transfer Level (RTL) hardware. We end the paper by applying the **eNDA**-flow through Syn2Logic to show how to — without writing a single line of hardware description language (HDL) code— **(i)** generate what we believe is the fastest *C. elegans* accelerator that runs significantly faster than state-of-the-art simulators, **(ii)** create (to the best of our knowledge) the fastest, most generic neuromorphic Sudoku solver that outperforms CP-SAT and SCIP, **(iii)** create a 5.6 million FPS/Watt accelerator on a tiny FPGA that outperforms existing neuromorphic architectures in terms of speed and energy-efficiency on the MNIST dataset, and **(iv)** create a generic accelerator for unsupervised learning that outperforms most existing unsupervised methods on the OPS-SAT benchmark.

# 1 Introduction

With the end of Dennard's scaling [1] and the impending end of Moore's law [2], the computer science community is actively pursuing alternative forms of computing to continue the performance scaling [3] we have grown accustomed to. Many of these alternative forms of computing are so-called post-Moore technologies, ranging from analog computers to quantum computers. However, none is perhaps as salient as that called neuromorphic computing.

Neuromorphic computing [4] is a computing paradigm that involves mapping the algorithm to a set of biologically inspired neurons and synapses, whose configuration solves computationally demanding tasks. Neurons can be viewed as actors that sparsely communicate with each other, while synapses are the strength of information between neurons. Neuromorphic computing is most often (but not always) implemented using Spiking Neural Networks (SNNs) [5]– the third generation neural network models. Unlike an imperative application running on a classical (Von-Neumann) system, neuromorphic systems can be both faster and more energy-efficient. Examples of the types of problems solved using neuromorphic computing include artificial intelligence (AI) [6–8], computational neuroscience [9, 10], graph traversal/processing [11], and combinatorial optimization problems [12–14].

Neuromorphic computing is often performed on a neuromorphic architecture [15, 16], which is a specialized architecture – either digital-, analog-, or mixed-signal – that emulates neurons and synapses *in silico*, exploiting the large amount of parallelism, event-driven nature, and sparseness that neuromorphic computing offers. To be called a neuromorphic architecture, it must [17] **(i)** compute using neurons and synapses, and **(ii)** communicate using spikes (action potentials); other traits such as being massively parallel, co-locating memory and computation, and being event-driven are important implementation design-decisions [16]. Some notable neuromorphic architectures include SpiNNaker I and II [18, 19], IBM TrueNorth [20], Intel Loihi [21], and BrainScaleS [22] [1].

Neuromorphic computing algorithms *require* neuromorphic architectures, as these algorithms do not map well to classical imperative computers. For example[2], **(i)** simulating one $mm^2$ cortical sheet [25] is 2.7x faster and 18.8x more energy-efficient on a neuromorphic Field-Programmable Gate Array (FPGA) than using an NVIDIA V100 GPU [26], **(ii)** solving a Least Absolute Shrinkage and Selection Operator (LASSO) optimization on Intel Loihi [23] is five orders of magnitude faster than on a traditional Intel CPU, and **(iii)** Deep-Learning (DL) inference on CIFAR10 [27] on IBM TrueNorth is 90x more energy-efficient than NVIDIA Titan X. Hence, to advance neuromorphic computing research and adoption, there is a fundamental *need* for high-performance neuromorphic architectures.

In this paper, we propose what future **e**lectronic **N**euromorphic **D**esign **A**utomation (**eNDA**) should be– a concept that extends the ideas of electronic design automation (EDA) *and* computational neuroscience to allow push-button generation of custom neuromorphic architectures from first-principles descriptions of neural dynamics targeting Field-Programmable Gate Arrays (FPGAs) and Application-Specific Integrated Circuits (ASICs). Additionally, we introduce Syn2Logic, which is a prototype implementation of the **eNDA** methodology concepts, including a custom, multi-paradigm, domain-specific language (DSL) that combines concepts of hardware description languages (e.g. VHDL and Verilog) with classical neural modeling language (e.g., NestML [28]) while also integrating templatized multiple dispatch-like (similar to Julia [29]), and methodologies for building neuromorphic neural networks. Finally, we introduce a class-4 [16] capable Syn2Logic backend that targets small-/mid-sized neuromorphic systems, which we use to perform **eNDA** exploration of neuron- and synapse-model and their networks, some of which have never been mapped down to silicon. We end the paper by pushing the state-of-the-art in neuromorphic systems in four ways: **(i)** we show how the proposed **eNDA**-flow can be used to create an accelerator that simulates the *Caenorhabditis elegans* (a nematode commonly used as a model organism) level-A (leaky integrate-and-fire) connectome faster than any other methodology, reaching between 392x-143,971x faster simulation speed than state-of-the-art simulators running on server-class systems, **(ii)** we develop a neural optimization solver for Sudoku, solving a 46-puzzle benchmark (including the notoriously hard AI escargot) up to 96.2x (ASIC estimates) and 28x (FPGA) faster than modern solvers such as SCIP [30] and CP-SAT [31], generalizes to other puzzles (including the famous top1465 puzzle set) and – for a few puzzles – is able to compete with tdoku[32]; the world's fastest, hand-written, SIMD-based Sudoku solver, **(iii)** we use the proposed **eNDA**-flow to show how to design a MNIST classifier that – to the best of our knowledge – is currently the fastest SNN MNIST classifier in existence, reaching 5.6 million frames-per-second/Watt and 1.72 million frames-per-second, outperforming existing FPGA SNN accelerators by several orders of magnitude and being faster than ASICs such as Intel Loihi and IBM TrueNorth using a tiny FPGA (Altera MAX10) from 2014, and **(iv)** we create a *generic* hierarchical neuromorphic accelerator supporting online unsupervised learning that performs second best out of 24 unsupervised methods on real space-telemetry OPS-SAT benchmark.

## 2 Neuromorphic Design Automation

Today, computational neuroscientists, neuromorphic algorithm developers, and neuromorphic hardware designers operate mostly in two isolated environments (Figure 1).

Computational neuroscientists adopt a flow (Figure 1:a) where algorithms and neuron- and synapse-dynamics are described using the very foundational mathematical models that govern them in computer simulator frameworks such as NEST [33], Brian2 [34], or GeNN [35], where the software is also responsible for executing the described models on some classical computer hardware, such as central processing units (CPUs), graphics processing units (GPUs), or the odd neuromorphic architecture (e.g., SpiNNaker [18]) that might support the model. Neuromorphic algorithm developers and end-users, similarly, leverage frameworks such as snnTorch [36] and SpikingJelly [37] for creating neural networks solving different problems. The observation to be made here is that these domain- and application-experts rely on general-purpose hardware for running their experiments– hardware that we know [23, 24, 26, 38] is not a good match for the event-driven and extremely parallel nature of said models. There is also a surge in programming frameworks such as Lava [39], NxTF [40], PyNN [41], Nengo [42], and Fugu [43] that attempt to bridge programming and mapping neuromorphic devices from a higher abstraction level to simplify the use of these devices. Proposals such as Neuromorphic Intermediate Representation (NIR) [44] provide an intermediate language that facilitates backend mapping to both neuromorphic and classical systems.

[1] For an exhaustive list, we redirect the reader to two excellent surveys on the topic [15, 16]

[2] For a more exhaustive list of opportunities, see the excellent summaries by Davies et al. [23], Schuman et al. [17], and Kudithipudi et al.[24].

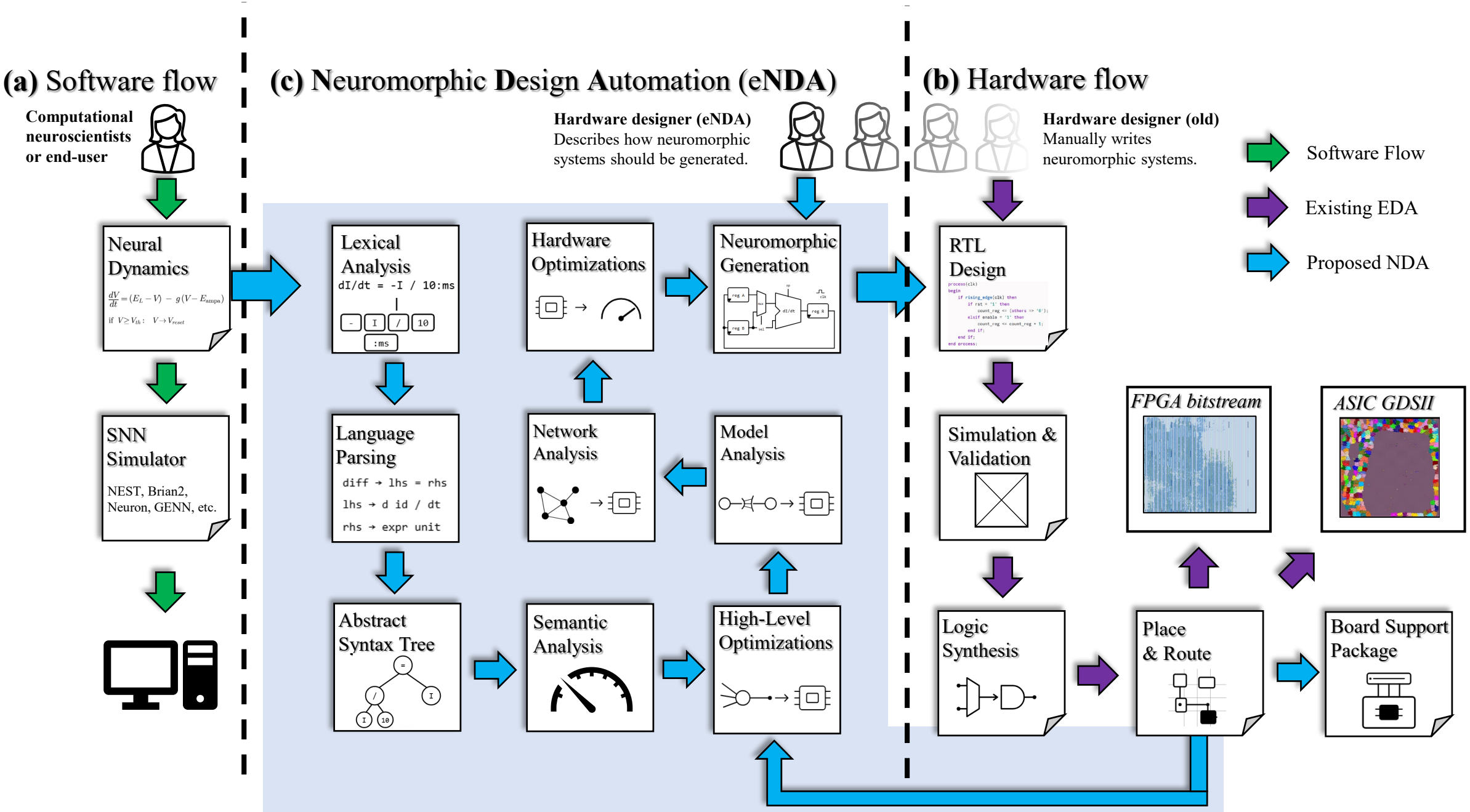

**Fig. 1**: (c) **eNDA** is the missing link between traditional neuroscientific algorithm development (a) and the well-known electronic design automation (EDA) flow that hardware designers use for ASIC/FPGA creation (b), allowing push-button generation of neuromorphic systems supporting state-of-the-art algorithms and increased technology transfer in the neuromorphic domain. **eNDA** shifts the role of hardware designers from recurrent manual and laborious hardware design to non-recurrent decision-making inside the framework, focusing on *how* a neuromorphic platform should be implemented.

A hardware designer, on the other hand, works with a different flow (Figure 1:b), where they decide on a particular neuron- and synapse-model and make several hardware-friendly optimizations to the model that may or may not be appropriate for the experiments performed by the domain experts. The final design is described using hardware description languages (VHDL, Verilog, or alternatives) and manually (often iteratively) mapped down towards either FPGA or ASIC using existing Electronic Design Automation (EDA) tools, resulting in an FPGA bitstream or ASIC GDSII output. There are two observations to be made here: **(i)** the entire process is very labor-intensive, and would have to be repeated for even the smallest changes made to the chosen models with no obvious ways of design-space exploration (DSE) and automation, **(ii)** the flow is often detached from the models and expectations that domain experts or end users actually need [3]. A prime example of the second observation is that it took more than a decade for Spike-Timing-Dependent Plasticity (STDP) [45] to make it into hardware after its discovery in 1997, and the model organism *C. elegans* [46], our largest fully mapped connectome (prior to the recently released Drosophila melanogaster [47]) has not been fully accelerated in hardware. And while there are several excellent SNN-to-FPGA/ASIC generation frameworks (see our recent survey on the topic [16]), they create **fixed-functionality** neuromorphic hardware based on architecture templates with fixed topology, parameters, and neuron/synapse model (often one or several IF/LIF variants), and do not offer the freedom of model/network expressivity that might be needed to develop new theories, use-cases, advanced models [4], and networks. Examples of such excellent work include DeepFire2 (S-CNN) [48], RANC [49] (parameterizable overlay), Spiker+ (LIF-based generator) [50], QUANTISENC [51], and HLS4ML [52] (that uses High-Level Synthesis).

However, to the best of our knowledge, no prior work combines **(i)** a hardware-oriented flexible domain-specific language for arbitrary models (from simple LIF to complex Hodgkin-Huxley and beyond), synapses, and plasticity dynamics, **(ii)** automated precision/bandwidth/timestep design-space exploration, and **(iii)** direct synthesis to both FPGA bitstreams and ASIC layouts.

[3] Worth mentioning is the Intel Loihi program that engaged the community in providing feedback for their platform in between generations.
[4] As the reader soon will see with our Sudoku-solver, even heavy modifications to the neural dynamics are sometimes needed to solve problems better in a neuromorphic architecture.

We propose the electronic Neuromorphic Design Automation (**eNDA**) methodology, shown in Figure 1:c, which bridges algorithmic and neuroscientific descriptions with EDA tool-flow, allowing application, models, and hardware designers to operate within one umbrella. An **eNDA**-flow takes as input the domain-specific model of a neural system as specified by a domain expert and automatically – through a series of complex transformations – transforms this high-level abstraction down to a custom and specialized neuromorphic platform, where EDA-flow takes it down to hardware. **eNDA** changes the hardware designer's role significantly. Rather than relying on insufficiently informed functional decisions regarding *what* should be implemented, designers now can focus on *how* a system should be implemented given a certain implementation strategy.

This view of **eNDA** opens up several opportunities. Firstly, it allows domain experts and hardware designers to work within a single unified framework, allowing improvements on one side to be directly transferred to another. For example, a new learning rule or network structure can directly be synthesized and made into a neuromorphic accelerator to solve problems in society. Secondly, the hardware designer now focuses exclusively on how hardware should be designed with direct access to a large number of domain-specific use-cases, allowing hardware decisions to be empirically explored systematically. Thirdly, domain experts can run their algorithms on more energy-efficient and faster FPGAs or even construct ASICs with the push of a button.

The flow in an **eNDA** can differ, but overall has the following functionality:

- A Lexical analysis and language parsing step for understanding the neural algorithms to be incepted as a neuromorphic accelerator (these could rely on descriptions written in NEST, Brian2, or even NIR),
- An intermediate representation (e.g., AST) encapsulating model dynamics and connectome and semantic analysis on said representation,
- A series of high-level, hardware-agnostic optimizations, such as (for example) constant folding or common sub-expression elimination,
- A model analysis step that understands and decorates the constructs with information critical for performance generation, such as whether a particular model is event-driven or continuous.
- A network analysis step that analyzes and decorates the network with information such as information flow, bandwidth requirements, etc,
- A hardware optimization step that performs hardware-specific optimization specific to the target, and
- Neuromorphic synthesis, where a *backend* generates the neuromorphic platform based on the decorations provided by the other parts.

Note that an **eNDA**-flow is not *tied* to a digital-only implementation, and different *backend*s can map the neuroscientific descriptions to digital, analogue, or even mixed-signal components subject only to the availability of such a backend.

## 2.1 Syn2Logic: an eNDA prototype

Syn2Logic is our prototype framework embodying the spirit of **eNDA**, implementing the flow of Figure 1. Syn2Logic operates under a custom-designed, multi-paradigm language that combines (and is inspired by) elements of traditional hardware description languages such as VHDL (concurrent, strictly typed, hierarchical), with neural modelling languages such as NestML [28] and Brian2 [34], and a variant of Julia's [29] multiple-dispatch functionality to effectively describe and build hierarchical neural networks. The language is LL(1) and object-oriented, where models and networks described can act as extendable prototypes, reducing code bloat and increasing readability and code reusability. In short, Syn2Logic combines the benefits of a hardware description language with the familiarity of neural description syntax.

The Syn2Logic model of computation has been created to be as compatible with existing frameworks as possible, simplifying the transfer of models, while still being hardware-friendly in order to facilitate mapping of descriptions to various types of hardware, such as digital and analog. The model of computation is divided into two types of components: a *model* and a *net*work.

A *model* is the smallest unit that can perform computation, and is composed of *state*, *update*, *event*s, and *conduits* (input) and *outputs*. A model consumes (and computes) data from conduits and drives outputs. Model state consists of variables and constants. The model update describes how to go from time `t` to time `t+1`, most often (but not exclusively) using Ordinary Differential Equations (ODEs). Model events are always-on, timeless monitors of the model state that trigger when a particular condition is reached. Generating output (that is, a spike) can only be done through events.

The second part of the model of computation is a *network*, consisting of one or more models or subnetworks connected to each other. Here, the user describes *how* a network is connected and how data is processed before entering each model. Each connection is event-driven, carries a token, and information is transferred *instantly*. Furthermore, connections occurring at conduits can be processed, such as, for example, accumulating all incoming data, before sending it to a model/subnet, subject to how the receivers require the data. Information from outputs to conduits occurs instantaneously (unlike models whose state is synchronously updated), and combinatorial paths can form in a network (akin to their digital design counterparts).

### 2.1.1 Model Objects

***Model Definition***

A generic model is defined using the keyword `model` followed by optional prototype information, then state, and a body of ODEs and events. When a prototype statement is provided, the model inherits state and connectivity rules (see network below) from the prototype, promoting code reuse and versatility.

Listing 1: A base-class model of a neuron with an empty body but with an interface, followed by another neuron that uses Neuron as a prototype.

```
model Neuron
 conduit Iin:pA
 conduit spike_in:mV
 output Vout:mV
 var      Vm:mV = -65:mV
 const Iconst:pA = 0,Vrst:mV = -65
begin
end

model LIF = Neuron(Vrst=-55:mV)
begin
...
end
```

***Model State***

State of the model contains `variables` or `constants`. A constant is a compile-time fixed value that cannot be changed after the design of the neuromorphic architecture, while a variable is subject to change as a function of time (and available for observation from outside the circuit). Both constants and variables can be typed, where a type is either a base (or derived) unit from the International System of Units (SI), a scalar value, or a custom type. Metric prefixes can be used to specify the magnitude of the unit (μ, M, G, etc.). The language is agnostic about the numerical size (in bits) of the variables or constants, leaving that decision to the compiler (or user). This allows variables to be size-*agnostic*, unlike (say) a language such as C or C++. All types of states can be made into an array or population. Modifying a variable can be performed in one of three ways: **(i)** externally by an end-user (when the system is not running), **(ii)** through time-dependent ordinary differential equations (ODEs), or **(iii)** imperatively inside events.

Listing 2: A constant (scalar) and a variable (nanosiemens)

```
const Aplus = 0.2
var chem_GABA:nS = 3
```

***Model Interfaces***

Interfaces are a special type of variable that are visible, have a sense direction (input or output), and are accessible outside of the model. They map very naturally to the neuromorphic concepts, since a neuron (or synapses, axons, astrocytes, glia, etc.) also shares the concepts of an input or output interface. There are two types of interfaces: a `conduit` and `output`. An output is an interface that can only be read from outside the model and is used by the model to transmit information out. An output in Syn2Logic is akin to an output port in (for example) VHDL, and may have only a single driver (avoiding race conditions). A conduit is an interface that can only be written into from outside the model and is used to transmit information into the model. Unlike an output, a conduit can have multiple writers to it, and it is up to the model to describe how to perceive those writes using collectives such as `sum`, `product`, or `max`.

Listing 3: A conduit (picoampere) and an output (scalar)

```
conduit I_in:pA
output Vout
```

#### *Model ODEs*

ODEs describe how variables in the model change as a function of time, and map naturally to the dynamics of neurons and synapses. An ODE block, follows textbook conventions ($\frac{dX}{dt} = ....$), allowing the code to be readable and relatable. ODEs at the language level have no time discretization; albeit (for digital hardware), hints regarding what kind of solver or timestep is needed for stable integration can be provided. All equations in Syn2Logic must match; the type of the left-hand expression must match that of the right-hand expression. Note that the magnitude (e.g., the resolution of 1: mV + 1: GV) may differ, and our compiler will insert required adapters between expressions at a later stage in the pipeline.

Listing 4: A differential equation for LIF neuron membrane dynamics

```
1 dVm/dt = (Iconst + Gexc*(Eex-Vm) + Ginh*(Ein-Vm) - gL*(Vm-EL))/Cm
```

#### *Model Events*

Events are anonymous functions that *can* modify model state asynchronously and atomically (appearing, to an observer, instantaneously), but can only be triggered upon a condition. A condition could be, for example, when a model of a neuron receives an action potential or spike from other neurons. Events that are triggered only as a function of external conduits are asynchronous (and can occur without advancing time), while events that are a function of at least one state variable are called synchronous (and can occur only during state update). Asynchronous events are important as they enable (for example) lazy updating of model state and pass-through (combinatorial) relay of signals (an event based on a conduit that writes to an output relays information downstream instantaneously).

Listing 5: An event that activates if internal state Vm>-30 and writes to output port Vout while resetting Vm to -60:mV, and an event that triggers when a conduit receives spikes, and accumulates the value of all spikes (::sum) on the conduit into variable Hinh

```
1 event(Vm > -30:mV)
2    Vout = Vm, Vm = -60:mV
3  end
4
5 event(I_in)
6     Hinh = Hinh + I_in::sum
7 end
```

### 2.1.2 Network Objects

#### *Network Object*

A `network` object in Syn2Logic acts as the blueprint for how to wire together populations of models (or other nets) to create a larger, more capable, component. Often, the blueprint of a neural network dictates its functionality, and we see many such examples from the literature, including central pattern generators (CPGs) [53], winner-take-all (WTA) [54], and even full-brain connectomes of invertebrates such as the nematode *C. elegans* [55] and the fruit fly *Drosophila melanogaster* [56]. Information between network objects is sent instantaneously from sender (output) to receiver (conduit).

Describing a network object involves three things: **(i)** instantiate population instances of models or networks and their configuration (e.g., changing constants). Each instance gets its own copy of the model state, and it is possible to override any value at instantiation (e.g., giving one neuron stronger drive than others), **(ii)** describe how populations wire together using well-known connectivity schemes such as *all2all* or *one2one*, connecting outputs of some models to conduits of others, and **(iii)** a network can (as a model) have their own conduits and outputs connecting internal populations to propagate information in and out of the network.

Listing 6: A simple WTA network that instantiates NZ neurons that inhibit each other and compete for activation.

```
1 net WTA
2   const NZ = 3
3   conduit cinp[NZ]:mV
4   output  cout[NZ]:mV
5   obj     pop[NZ]:LIF()
6 begin
7   one2one(cinp[*]      ⇒ pop[*].spike_in)
8   one2one(pop[*].Vout ⇒ cout[*])
9   one2all(pop[*].Vout ⇒ pop[*].spike_in)
10 end
```

***Rule-based Connectivity***

Connecting outputs of one model to the input of another is trivial as long as the types match, such as connecting an output of type $mV$ to an input also of type $mV$. However, this quickly becomes very verbose and unclean code-wise as it combines *how* something is connected with the *topology*. In Syn2Logic, we decouple how something is connected from the network itself by using **rule-based** connectivity. A `rule` describes how a `model` or `net` connects to other components in the system. For example, a simple synapse would connect its output (conductance) to a neuron's conduit, and a more complex plastic synapse would also connect the post-synaptic neuron's output back to itself. Furthermore, rules are inherited through prototypes, meaning that if a rule connecting our basic empty `neuron` (defined above) with a basic empty `synapse` exists, then all variants of neurons and synapses inherited from their respective prototype will know how to connect unless a new rule for that model/net is written. The concept is similar to Julia's multiple dispatch.

Listing 7: A rule describing how a WTA network connects to another WTA network, in this case by having each entry in the connecting WTA inhibit corresponding entry in the destination WTA. Such a network is useful for when, for example, creating neuromorphic constraint satisfaction solvers.

```
1 rule src:WTA ⇒ dst:WTA
2 begin
3   one2one(src.cout[*] ⇒ dst.cinp[*])
4 end
```

***Template-based Instantiation***

Finally, by decoupling *what* connects to what from *how* it connects, we can allow templates. A template is a construct that allows a network to operate on generic models/networks, which are resolved at compile time and whose connections use the ruleset that has been defined. Templating allows writing generic code for constructing a network without specifying what type of neuron or synapse model is being used. For example, one can define a network describing a microcircuit using a template and then easily change the underlying neuron model from LIF to Hodgkin-Huxley, or even to other sub-networks with the change of a single line (or during instantiation).

Listing 8: A simple two-neuron oscillator built using templates (`TNeuron` defaults to LIF, `TSynapse` defaults to AMPA) allowing any template to be changed to any other model (or network) assuming the rules for connecting them exist.

```
1  net HCO
2    template TNeuron = LIF()
3    template TSynapse = AMPA()
4    obj neu[2]:TNeuron()
5    obj syn[2]:TSynapse()
6    output spkout[2]:mV
7  begin
8    one2one(neu[0] ⇒ syn[0])
9    one2one(syn[0] ⇒ neu[1])
10   one2one(neu[1] ⇒ syn[1])
11   one2one(syn[1] ⇒ neu[0])
12   one2one(neu[*] ⇒ spkout[*])
13 end
```

## 2.2 Example Network

To demonstrate a complete example, consider Figure 2:a, where we see a network not entirely unlike the nervous system for the *C. elegans* nematode, ported to the Syn2Logic language from c302 [55]. Red, blue, and green nodes correspond to different neuron types (inter-, motor-, and sensory-neuron), and gray and white boxes correspond to plastic and non-plastic synapses, respectively. Our example network works with STDP synapses, which c302 does not, and this is only to show the different perspective of the Syn2Logic language (we construct a more faithful c302-variant later in Section3.3).

Figure 2:b shows an enlarged part of the graph, showing a motor neuron (red) and its description in the Syn2Logic language (Figure 2:c). A plastic STDP synapse (Figure 2:d) is connected to the motor neuron using the specified ruleset (Figure 2:e) by connecting the output conductance $G_out$ to the $chem_in$ conduit of the neuron, and connecting the spiking output $Vout$ of the neuron back to the $Vpost$ conduit terminal of the synapse, completing the connection. A simpler non-plastic synapse is shown in Figure 2:f.

Parts of the top-level *C. elegans* -like Syn2Logic description are shown in Figure 2:a, where the neuron and synapse objects are created followed by a long connectivity list. This is a detailed and verbose way of describing each connection in the system. An alternative – and more general – approach is shown in Figure 2:g, where we create a generic accelerator with 302 neurons and $302^2$ STDP and inhibitory synapses, respectively, and connect all of them together.

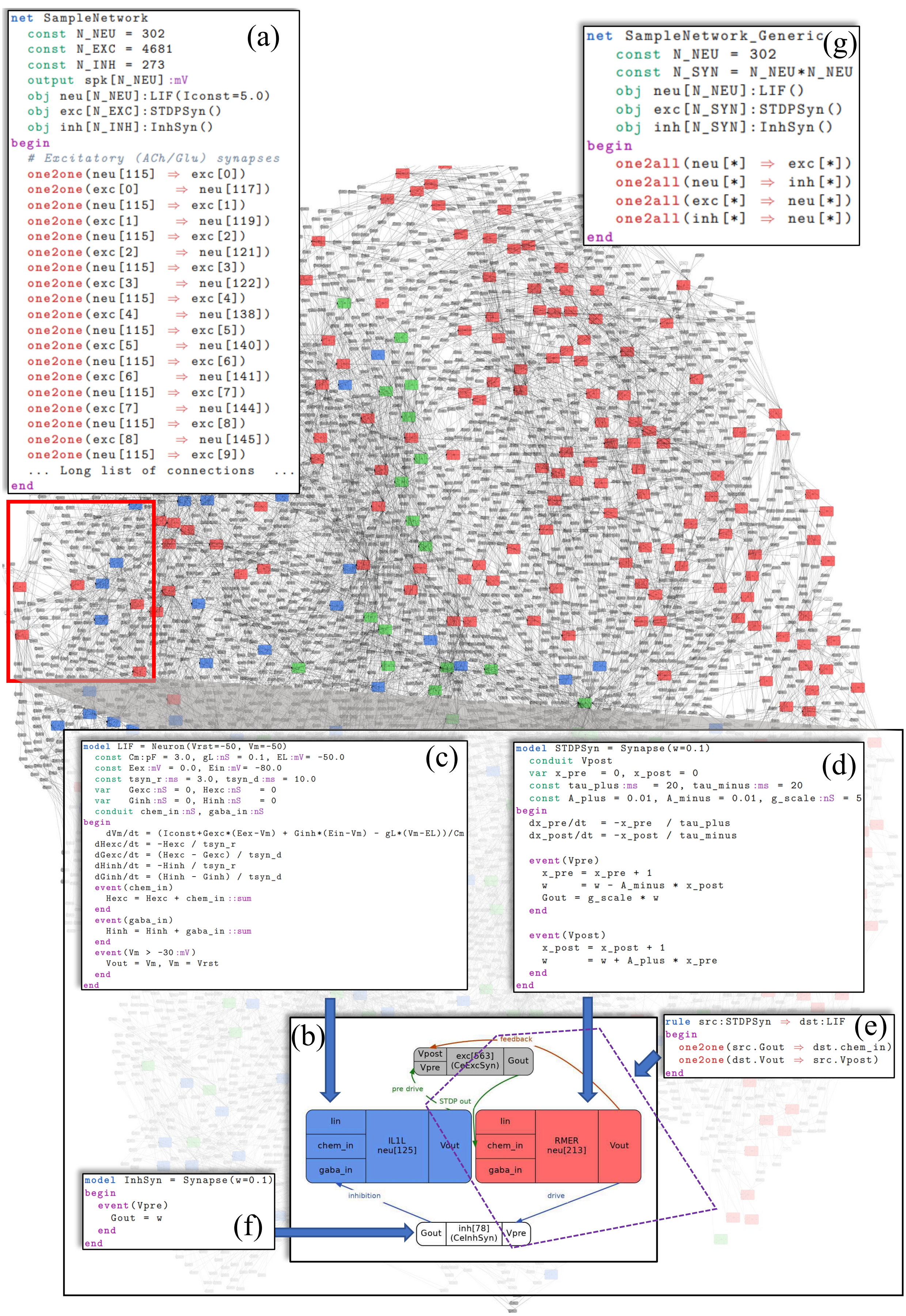

**Fig. 2**: A Syn2Logic graph (rendered using dot and based on the *C. elegans* nervous system), showing the top-level Syn2Logic code describing it (**a**), a zoom-in of a motif (**b**) where we see the neuron model (**c**), the STDP synapse (**d**), and the rule (**e**) on how to connect an STDP synapse to the neuron, which is also visible in the motif. We also see the simpler non-plastic synapse (**f**). Finally, we see a generic description of the accelerator (**g**) that is condensed but requires the user to configure it (unlike **a**, whose connectivity is sparser but immutable).

Smaller and more elegant, this description requires the users to actively program the state of the accelerator prior to execution (similar to how neuromorphic systems such as Intel Loihi are programmed).

## 2.3 Syn2Logic Neuromorphic Backend

As with a regular compiler, Syn2Logic can have multiple backends transforming the intermediate representation into a digital, analog, or mixed-signal neuromorphic device, each with its own characteristics, strengths, and weaknesses. For example, a neuromorphic device targeting a specific Internet-of-Things (IoT) use-case will likely use a backend focusing on creating an ASIC with low power consumption, while a generic design targeting a high-performance, high-capacity network would prefer a top-tier FPGA backend.

**Fig. 3**: A generic view of a model component generated by our Syn2Logic class-4 backend. We see an Izhikevich model with several state variables, and each variable is daisy-chained on the scan-chain, which goes serially through all components in the system and can be configured from outside. The model state is connected to a generated data path that solves the ODEs and handles events. Conduits are currents from (typically connected synapses) that are added together to provide reduction functionality at the neuron side. Each conduit and output has a token (a spike-enable) signal, indicating a valid value (not entirely unlike data-flow).

To illustrate the concept, we developed a backend that generates digital hardware that embraces the spirit of neuromorphic computing [57], where the substrate becomes the computation (the entire network is placed *in silico*, with no time multiplexing of data paths). The output of our backend is a component shown in Figure 3 (in this case, an Izhikevich neuron). The generated neuromorphic system meets the neuromorphic requirements described by Schuman et al. [17], and is considered a class-4 system per Szczerek and Podobas' taxonomy [16].

Our backend takes the intermediate representation and generates a digital neuromorphic system **(i)** whose components are discretely and spatially placed on the underlying silicon, **(ii)** communicates in a sparse and event-based fashion, **(iii)** implements data-paths for updating neural state using fixed-point representations, and **(iv)** is deterministic in that advancing one time step takes exactly one clock cycle. The strengths of our backend are that it generates high-performance yet deterministic neuromorphic systems without reliance on external peripherals (e.g., DDR4), and where it is therefore the closest to nervous systems in nature. The weakness is that it is not very scalable, and suitable only for small-to-medium sized designs.

Additionally, our backend accepts a number of optimization *knobs* with which the user can adjust or control the quality of the neuromorphic hardware generated:

- **Fixed-Point Width (M:N)**: A user can specify the size of state in terms of integer and fraction bits, which can trade accuracy for smaller silicon,
- **Timestep (TS)**: A user can select how fast time advances in the neuromorphic device. A large timestep is faster but can lead to instabilities,

- **Per-Model Bandwidth (PMW)**: A user can specify the bandwidth of conduits into a model, and can optimize for the fact that spikes are very sparse at the risk of dropping information in times of high activity.
- **Fused operations (FO)**: Forces fused data-paths and reduction adapters operating without loss in accuracy, and rounds only at the end of the computation,
- **Approximate Constants (AC)**: allows approximating constants to save silicon area, relying on the observation that some constants consume a lot of area to be exactly (if at all possible) defined.

## 2.4 Syn2Logic Complete Internal Flow

The internal Syn2Logic compilation flow follows our proposed **eNDA**-flow closely. To compile a Syn2Logic (`.s2l`) source file into a neuromorphic system in RTL, we perform the following steps:

- **Lexical analysis**: Our Flex-generated scanner deposits tokens onto a double-ended queue consumed by the recursive-descent parser. The lexer normalizes the language's notation: the equivalent derivative spellings `X'` and `dX/dt` are mapped to a single ODE token, SI units carry their metric prefix as a power-of-ten magnitude, and the identifier rule consults the symbol table *during scanning*, such that an identifier with a registered datatype suffix (e.g., `mV`) is emitted as a typed token.
- **Syntactic analysis:** Parsing is performed by a hand-written recursive-descent parser dispatching on the constructs `type`, `model`, `net`, `rule`, and `build`. Expressions are parsed with a two-stack operator-precedence scheme; integer exponents expand into repeated multiplications at parse time. Postfix analysis yields function calls, array accesses (including ranges and the * wildcard), member accesses, `::sum/::num/::max/::product` reductions, and `::type` casts. For models, the parser collects storage declarations (`const`, `var`, `conduit`, `output`) together with ODEs, algebraic equations, and `event` blocks; the left-hand side of an ODE is intentionally represented as $X/(1\,\text{ms})$ so that dimensional analysis of $dX/dt$ remains well defined. Nets comprise object and template declarations and connection commands, whereas rules are registered under the mangled name *srcType+dstType*. Prototype derivation (`model Foo = Neuron(...)`) clones the prototype's AST and overrides its constants from the supplied dictionary. Finally, `parse_build()` deep-clones the build target, applies its dictionary, and invokes **`Instantiate`**, **`AnalyzePorts`**, the optional visualization passes, and **`Class4`**.
  - **Verification**: This pass constitutes the principal semantic analysis and is re-executed for every instantiated clone. It resolves all symbols (and decorates the tree) and performs dimensional type checking over the SI unit system: addition, subtraction, comparison, assignment, and ODEs require identical dimensions on both sides, whereas multiplication and division combine dimensions algebraically; magnitude discrepancies merely warn, deferring to **`BalanceTypes`**. The pass further propagates information such as whether an expression is a constant, is writeable, or is event driven (a model writing its outputs solely in such handlers is itself classified as event-driven) and enforces the language's structural invariants: single-writer outputs, no variable/output/self reads in equations, reductions restricted to conduits (recording the `reduce_*` attributes consumed downstream), array dimensionality, function signatures, and constant, type-correct initialisers.
  - **Print (optional)**: A diagnostic pass that emits the fully attributed AST; its output is suppressed at default verbosity and exposed through command-line options.
  - **Instantiate**: This pass elaborates the declarative design into a concrete instance hierarchy: for a `Net`, it first applies the sub-sequence **`Verify`**→**`StripTypes`**→**`ConstFold`**→**`DivToMul`** (reducing array bounds to literals) and then, for each declared object, clones the referenced model or net, assigns a globally unique name to preclude duplicate VHDL entities, applies the parameter dictionary, attaches the clone via `setImplementation()`, and recurses depth-first. For a `Model`, the sub-sequence **`Verify`**→**`BalanceTypes`**→**`StripTypes`**→**`ConstFold`**→**`DivToMul`**→**`Verify`** is applied, after which the model consists of dimensionless, constant-folded arithmetic suitable for hardware mapping.
    - **BalanceTypes**: Source programs may freely mix magnitudes of a common dimension (e.g., millivolts and volts); for every binary statement, this pass compares the magnitude exponents of the operands and, where they differ, rewrites the right-hand side with an explicit multiplication or division by $10^k$, so that the subsequent erasure of units incurs no numerical loss.
    - **StripTypes**: This pass erases the unit system: cast nodes are replaced by their operand, the implicit $/\ 1$ ms on ODE left-hand sides is removed (right-hand-side divisions are preserved), and all declared symbol types are reset to the dimensionless type, leaving a purely numerical program.
    - **ConstFold**: Constant symbols are substituted by their values, operators over two literals are collapsed, and the identities $x+0$, $x-0$, $0-x=-x$, $x*0=0$, and $\frac{x}{1}=x$ are applied. As instantiation converts model parameters into constants, this pass effectively specialises each instance's equations and renders divisors literal for **`DivToMul`**.

* **DivToMul**: Since a hardware divider is substantially larger and slower than a multiplier, every division by a literal $c$ is strength-reduced to $X * \frac{1}{c}$, the reciprocal being evaluated at compile time.
* **Build command interpreter**: **`Build`** is the common base class of **`AnalyzePorts`** and **`Class4`**; subclasses override a virtual `connect()` method invoked once per resolved point-to-point link. Its `Command` visitor expands wildcards and ranges into index sets (verifying `one2one` cardinality) and enumerates index pairs for primitive endpoints. For object-to-object connections it performs rule expansion: the rule registered under *srcType*+*dstType* is cloned, its `src`/`dst` placeholders are substituted with concrete array accesses (**`ReplaceId`**), the result is re-verified, and the rule's commands are re-executed per index pair, unfolding a single net-level command into all of its port-level links.
* **AnalyzePorts**: The hardware required at a port depends on the opposite endpoint of each connection: a conduit read as `Iin::sum` obliges every connected source to export a value bus in addition to its spike bit. Inheriting the **`Build`** interpreter, this pass re-executes all connection commands, transfers each destination conduit's `reduce_*` attributes onto the corresponding source ports, and recurses through the instance hierarchy, thereby determining which reduction adaptors and `_val` side channels the backend must generate.

– **DotGraph/DotGraphPop**: Optional passes rendering Graphviz visualisations of the design at object and population level. The previous Figure 2 was generated using this pass.

- **Backend (Class4)**: The final backend produces the final RTL for the neuromorphic system, and is composed of four parts: expression mapping, differential equations/state mapping, model emission, and network emission:
  – **Expression Mapping**: Code generation proceeds bottom-up: each AST node deposits a `"mapped"` signal annotated with exact $(m, n)$ bit bounds. Literals are quantised to their minimal exact format, further reduced under `-round-constants`. Bit growth is tracked exactly (addition: max+1; multiplication: $m_l$+$m_r$+1; etc.), whereupon we resizes each result to $Q_{M:N}$; under `-fused-ops` the wide intermediate is instead retained, exchanging area for a datapath free of intermediate rounding. Comparisons yield single-bit `when/else` signals, and reduction nodes read the conduit's precomputed port. The functions `exp` and `log` are expanded inline into Taylor-series datapaths, the former via base-2 decomposition with a barrel shifter.

  – **ODEs, events, and state threading.** ODEs are integrated by the forward Euler method: $X + f(\ldots) \cdot TS$ also breaks is computed combinationally but not committed to the register, being published instead in a per-variable `NextState` map. Within an event, reads are redirected through `NextState`, so that a threshold condition observes the freshly integrated value ("integrate-then-test", as in Brian2/NEST) and an accumulating assignment composes with, rather than supersedes, the ODE step; each event assignment becomes a multiplexer over the value produced thus far, replacing the corresponding `NextState` entry; output assignments are emitted combinationally (event-driven models) or as guarded sequential statements. Only upon completion of the model visit is each variable's final value committed to its state (a part of `config` on the scan-chain) — one register write per variable per timestep, avoiding VHDL's last-assignment-wins semantics. Reads outside events continue to observe the registered value, preserving the simultaneous (Jacobi-style) integration of coupled ODEs.

  – **Model synthesis**: The backend produces a fully parallel implementation with co-located memory and computation (class-4 [16]): one VHDL entity per model and one structural entity per net, appended to `out.vhdl` together with a JSON device tree (`out.json`, which describes what variable is located where on the scan-chain). Its parameters ($M$:$N$ format, timestep `TS`, `pmw`, `round_constants`, `fused_operations`) come from the symbol table. Each `var` is realized as an `sfixed(M downto -N)` signal aliased onto a slice of a single `config` shift register, with its reset value, read alias, and bit address recorded in the device tree; the register simultaneously serves as a scan chain (`config_in/out/en`) for state load and read-back. Conduits map to spike input ports (with `_sum/_num`/etc. side ports where required), and outputs to output ports (with a `_val` bus where a downstream reduction consumes the value).

  – **Network synthesis**: For a `Net`, the backend first recurses into each object's implementation and then constructs a structural architecture. Re-executed `connect()` calls accumulate per-port lists of mangled signal names (`obj_id`*i*`_port_id`*j*); the instantiation loop then emits one `port map` per array element, in which spike inputs are combined by disjunction over all sources, `::num` is realised as an adder over spike bits, `::sum` as a balanced adder tree over `_val` buses (resized after each addition, of depth $\lceil \log_2 K \rceil$), and `::product` as a multiplication chain. For $0 < pmw < 1$, a reduction's $K$ inputs are partitioned into $\lceil K \cdot pmw \rceil$ bins of one priority-selected value each, trading connectivity for area. Per-model scan chains are concatenated through a `config_chain` signal; the architecture is appended to `out.vhdl` and the device tree written to `out.json`.

- **Post-Emission flows**: Following the backend, `parse_build()` invokes the flow generator selected by `-flow` — **`GenCadence`** (Genus+Innovus/ASAP7; default), **`GenOpenROAD`** (NanGate45, unused in this paper, but we do have preliminary support for OpenROAD [58]), or **`GenQuartusAgilex7`** — each bundled with `out.vhdl` into `<flow>_flow.tgz`; `-testbench` additionally generates a spike-counting **`GenMeasureOutActivity`** testbench.

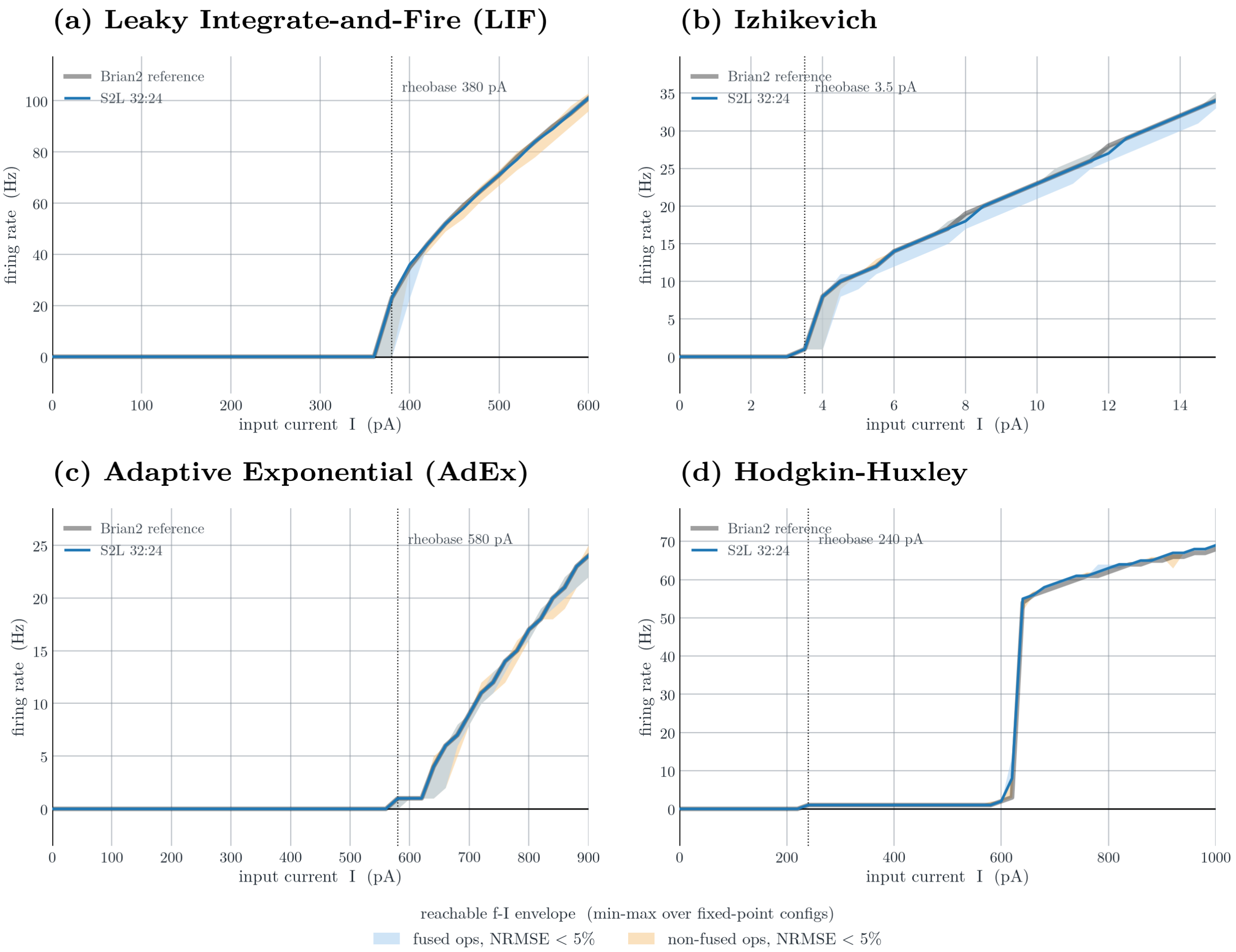

**Fig. 4**: Current-Frequency response (f-I) for four neuron models with reference framework Brian2 and Syn2Logic, including (a) Leaky Integrate-and-Fire, (b) Izhikevich, (c) AdEx, and (d) Hodgkin-Huxley.

# 3 Results

## 3.1 Neuron Correctness and Exploration

We start by exploring the correctness results for our proposed **eNDA**-flow for four different neuron models: Leaky Integrate and Fire (LIF), Izhikevich, Adaptive Exponential Integrate and Fire (AdEx), and the Hodgkin-Huxley model, in increasing order of complexity. For each evaluation, we see the firing rate of the neuron as a function of the internal constant current, a well-known correctness experiment. We compare to Brian2 as a golden reference, and use Normalized Root Mean Square Error (NRMSE) as a measurement for how well the different implementations match our Brian2 reference.

The LIF model – a cornerstone neuron model in neuroscience and neuromorphic engineering – is shown in Figure 4:a, where we see the f-I response curve for both the Brian2 reference implementation and the Syn2Logic $Q_{32:24}$ high-fidelity variant. Overall, this high-fidelity Syn2Logic variant faithfully follows Brian2's LIF dynamics, and – for the sampled input current points – yields the same 380 pA rheobase current, which is when (for our samples) the neuron starts firing. The error between this high-fidelity Syn2Logic LIF model and the Brian2 reference is as little as 0.4% NRMSE. The LIF model does not need an expensive $Q_{32:24}$ format to work well. In fact, by applying fused operations, we can operate with state variables at a $Q_{7:7}$ format with less than 5% NRMSE, and we are forced to widen the data-path to a $Q_{16:6}$ format if we disable fused operations and force-round results after each operation in the datapath. The NRMSE <5% requirement follows the trend of the Brian2 curve, albeit with some small over- and undershoots in the frequency for a given input current.

The Izhikevich model, which is more complex than the earlier LIF model, is shown in Figure 4:b. As with the LIF model, the Izhikevich Syn2Logic variant follows the Brian2 model faithfully, yielding identical rheobase currents for the tested current points and follows the curve near-exactly, with – interestingly – the non-fused version reaches 0.0% NRMSE error compared to Brian2, and the fused variant reaches 0.7% NRMSE. As with the LIF model, the Izhikevich model does not need the full $Q_{32:24}$ format, but – assuming an NRMSE <5% ratio can be tolerated – can work well with a $Q_{6:9}$ fused variant or a $Q_{13:8}$ non-fused variant, showing small under- and overshoots in the graph relative to the Brian2 reference value.

The AdEx model is even more complex than Izhikevich, containing more complex ODEs and also the hardware-unfriendly `exp` function. We see the comparison between Syn2Logic and Brian2 in Figure 4:c. The high-fidelity $Q_{32:24}$ Syn2Logic variant behaves identically with the Brian2 reference model, with a rheobase current of 580 pA. The fused-op variant stays at 0.0% NRMSE compared to Brian2 until we drop to 8 fractional bits, while the non-fused variant is a bit more chaotic. For fused-operations, a $Q_{9:6}$ value is enough to stay within NRMSE <5% (with notable undershoots compared to the Brian2 reference) while the non-fused variant reaches a 2% error rate at $Q_{16:7}$.

We end our validation experiments by comparing the Hodgkin-Huxley model on Brian2 and Syn2Logic, shown in Figure 4:d. The H-H model is a complex neuron model, with multiple ODEs and a large number of multiplications, divisions, and `exp` operations. While the Syn2Logic $Q_{32:24}$ variant reproduces and follows the Brian2 model rather well, it still has an NRMSE of 1.3% that cannot be reduced; interestingly, in fixed-point representation, a larger data type is not better, as $Q_{18:14}$ yields better NRMSE than $Q_{32:24}$ (for the non-fused version). This could be attributed to Brian2 model's IEEE 754 numerical representation being better than the fixed-point representation our backend operates on. The fused variant can be as small as $Q_{7:17}$ while keeping within a <5% NRMSE while the non-fused variant requires more integer bits and is at $Q_{16:14}$.

### 3.1.1 Discussion: Neuron Models in Neuromorphic Systems?

Synthesizing the neurons using Syn2Logic and Cadence Genus, we observe several trends, shown in Table 1. The first observation, which is rather intuitive, is that the more complex the neuron model, the more area it consumes and the slower it operates. For example, the LIF model can operate at 137 MHz, consuming less than 500 $\mu m^2$ in area, while the HH model, which is many times more complex, is significantly slower and can be more than 100x larger.

**Table 1**: ASIC cost of the four neuron models at their faithful (≤5%) fixed-point format, fused vs non-fused ops (ASAP7 7 nm, Cadence Genus estimation). *Area saving* and $\Delta F_{\max}$ are the non-fused variant relative to fused.

| Model | Ops | Area ($\mu$m$^2$) | $F_{\max}$ (MHz) | Area saving | $\Delta F_{\max}$ (MHz) |
|---|---|---|---|---|---|
| LIF | fused $Q_{7:7}$<br>non-f. Q16:6 | 475.6<br>439.4 | 137<br>129 | 7.6% | −8 |
| Izh. | fused Q6:9<br>non-f. $Q_{13:8}$ | 826.2<br>667.0 | 127<br>121 | 19.3% | −6 |
| AdEx | fused $Q_{9:6}$<br>non-f. $Q_{16:7}$ | 1020.0<br>783.7 | 99<br>65 | 23.2% | −34 |
| HH | fused $Q_{7:17}$<br>non-f. $Q_{16:14}$ | 52 272<br>13 212 | 30<br>30 | 74.7% | 0 |

The second observation is that fused data paths can give the neuromorphic system higher performance. Rounding fixed-point numbers can be expensive, and by rounding only when you store the result back, you can increase the operating frequency of the circuit. This increase is visible in LIF, Izhikevich, and AdEx, but is most notable in the latter (nearly 50% increase in clock frequency). However, this increase in clock frequency is not free, as the fused data paths are wider and thus – as a consequence – cost more area, whereas the non-fused version can have a footprint between 7.6% and 74.7% smaller than its fused counterpart.

An interesting observation is that, despite using a rather small, 7 nm, predictive PDK, our described neurons are still far away from their biological counterparts. Granule cells (among the smallest cells in the cerebellum, 8 $\mu m^2$) [59], cortical pyramidal cells ( 25 $\mu m^2$) [60], and the large Betz cells ( 100 $\mu m^2$) [61] are still 55x, 18x, and 4.4x smaller than even our most pruned and smallest LIF model, leaving ample future opportunities for improvements.

Finally, a word of warning: the exact width of the fixed-point representation is – as we will see in our use-cases – very use-case dependent. A computational neuroscientist might accept an NRMSE<5% (or stricter), or might opt for a larger data-path that yields better accuracy. A neuromorphic application, on the other hand, might actually work with a significantly smaller data-path– an observation we will see later when creating our neuromorphic Sudoku solver.

## 3.2 Synapse Correctness and Exploration

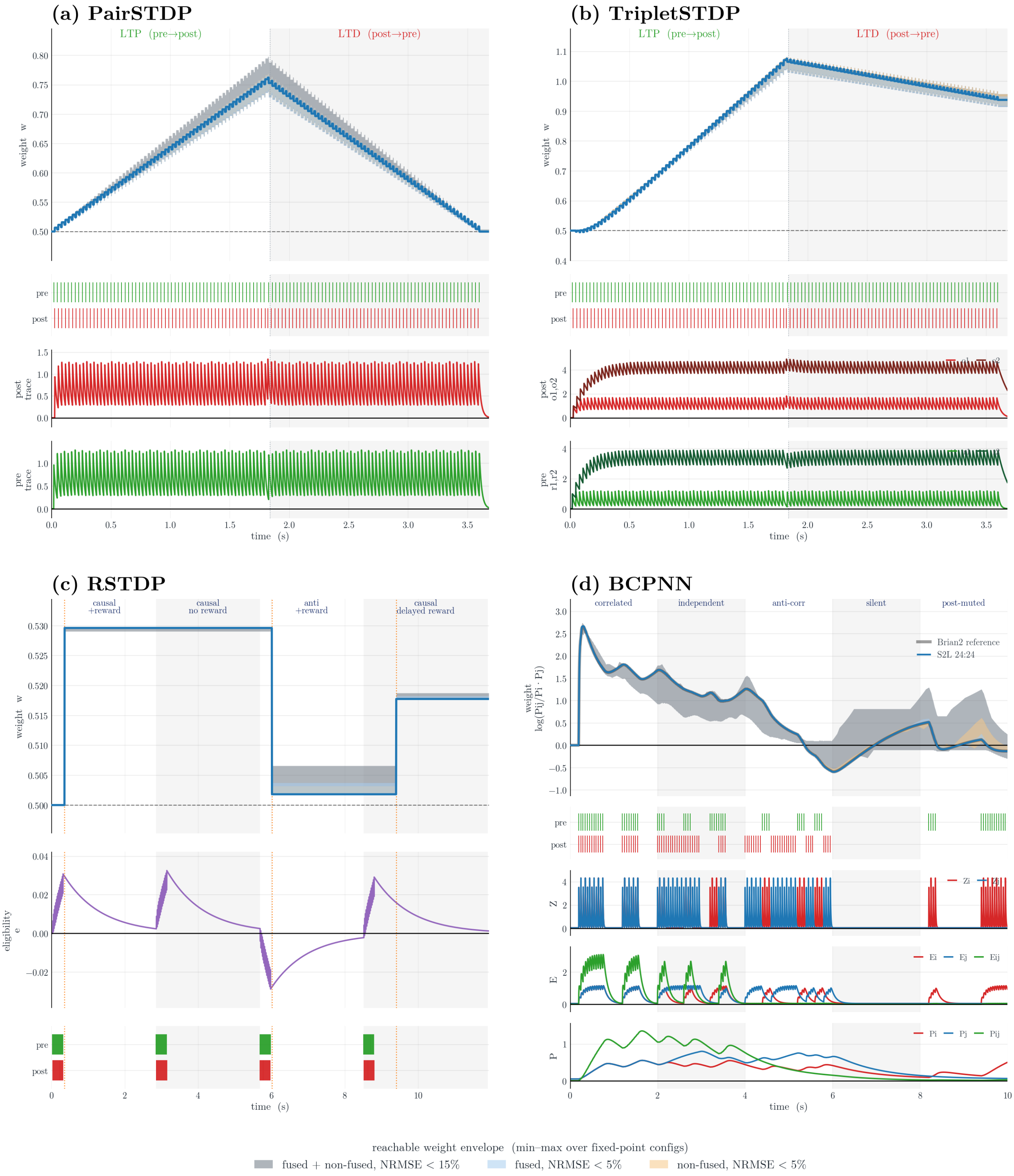

**Fig. 5**: Large-scale evaluation of four plastic synapse models (PairSTDP, TripletSTDP, RSTDP, and BCPNN) in the Syn2Logic framework, with the tested induction protocol (pre- and post-synaptical activity) and the change in weight (w) and other state variables. For each model we also show its sensitivity to fixed-point precision quantization.

Next, we quantitatively verify the correctness and explore the data-path pipeline for four well-known plastic synapse models and a specific induction protocol: Pairwise STDP, Triplet STDP, RSTDP, and BCPNN. The results are shown in Figure 5.

Figure 5:a shows the classical pairwise STDP rule based on traces. Here, we created an induction protocol that first encourages synapse potentiation (the weight w increases) for the first half-second by driving post-synaptic spikes directly after pre-synaptic spikes, and then induces synapse depression (the weight w decreases) during the remaining half-second of the simulation. The functionality, the traces, and the depression/potentiation are in line with the expected functionality of STDP. Furthermore, what is interesting is that STDP, as long as the weights are clipped to a maximum, can tolerate a data-path width reduction as low as $Q_{2:8}$, yielding an NRMSE of 3.5% and 4.1% for fused- and non-fused data paths alike; going further significantly decreases the performance of the learning rule compared to its golden reference.

Figure 5:b shows the TripletSTDP rule with a similar induction protocol as the Pairwise rule, but containing twice as many trace variables. The rule functions as expected in both depression and potentiation modes, and — due to the more complex nature of the rule — requires more fractional bits compared to its pairwise counterpart, landing at $Q_{2:9}$ to maintain 2.8% and 3.5% for the fused- and non-fused variants' NRMSE, respectively, over the golden reference.

Figure 5:c shows the reward-modulated STDP rule (RSTDP), with a more complex induction protocol consisting of both causal and delayed rewards, as well as no rewards and anti-rewards. We see how the different pre- and post-traces induce changes to the synapse strength based on the eligibility trace for the synapse weight. Our chosen induction protocol actually requires quite many fractional bits to work as intended, with both the non-fused and fused variants requiring a $Q_{2:20}$ format, yielding 2.49% and 3.50% NRMSE in favor of the non-fused variant.

Finally, we take on a much more complex learning rule in Figure 5:d, where we see the Bayesian Confidence Propagation Neural Network (BCPNN) learning rule. Here, we use a complex induction protocol that includes five different activities on the eligibility traces (correlated, independent, anti-correlated, silent, and post-muted). In addition, we compare our Syn2Logic-generated functionality against the reference Brian2 model. At a $Q_{24:24}$ data-path, our Syn2Logic variant mimics the Brian2 model almost exactly in all regimes. In fact, we can reduce the data-path width to as little as a $Q_{8:18}$ (non-fused) and $Q_{8:19}$ (fused) width while still being 3.5% and 0.92% NRMSE compared to the $Q_{24:24}$ variant.

In the end, whether a plasticity rule works or not is likely scenario-dependent, and could allow for further reductions than shown here. For example, if an NRMSE<15% error rate can be tolerated (the gray shadow behind the graph in Figure 5:d), then $Q_{8:14}$ format should suffice. Syn2Logic can and will be used in the future to quantify the impact of plasticity rules and for more complex use-cases.

### 3.2.1 Discussion: Which Synapse Model to use in Neuromorphic Systems?

The synthesis results for the synapses are shown in Table 2. As expected, the Pairwise-STDP synapse is by far the smallest, consuming as little as 60.4 $\mu m^2$ area, with the Triplet-STDP (larger fused variant) consuming between 105 and 374 $\mu m^2$, RSTDP between 538 and 657 $\mu m^2$, and BCPNN – being the most complex and using a `log` function – is the most expensive at 1386.9 $\mu m^2$. As expected, the fused variants are larger than their non-fused counterparts, with the highest benefit observed in TripletSTDP (71.7%). Interestingly, the BCPNN $Q_{8:19}$ version synthesized was smaller than the non-fused version.

**Table 2**: ASIC area of the four plasticity rules at their faithful (≤5%) fixed-point format, fused vs non-fused ops (ASAP7 7 nm, Genus synthesis estimations). *Area saving* is the non-fused variant relative to fused.

| Rule | Ops | Area ($\mu$m$^2$) | Area saving |
|---|---|---|---|
| Pair | fused $Q_{2:8}$<br>non-f. $Q_{2:8}$ | 91.9<br>60.4 | 34.3% |
| Triplet | fused $Q_{2:9}$<br>non-f. $Q_{2:9}$ | 374.4<br>105.9 | 71.7% |
| RSTDP | fused $Q_{2:20}$<br>non-f. $Q_{2:20}$ | 657.1<br>538.3 | 18.1% |
| BCPNN | fused $Q_{8:19}$<br>non-f. $Q_{8:18}$ | 1 269.9<br>1 386.9 | −9.2% |

While the difference in hardware cost is – as expected – different in the various plastic synapses, it does raise another question: how much "learning" capacity does a synapse actually bring for the silicon footprint it consumes? If a neuromorphic IP component has a silicon budget of 0.1 $mm^2$ for a device that should learn online using plastic synapses, should that silicon be spent on 78 BCPNN synapses, 185 RSTDP synapses, 944 TripletSTDP synapses, or 1655 STDP synapses to perform the best? Likely, the answer depends on the use-case, where some use cases might benefit from the more advanced learning of BCPNN synapses while others favor the simpler (but more) Pair-STDP synapses. Independent of which, the amount of "learning" per silicon area is seldom discussed, but should be quantified in the future, since – in the end – it is the hardware implementation that in part dictates the success (through lowered execution time or reduced power consumption) of putting neuromorphic systems to good use. An example metric could be Silicon Learning Fidelity, $SLF = \frac{\#Synapses}{mm^2}$ for a particular technology node and iso-fidelity.

We end the discussion by comparing our results to biology [5]. Unlike neurons, which *could* (the smallest LIF model) be between 4x-55x larger than biological neurons, the same – unfortunately – does not hold for synapses. A typical synapse in the human brain is on the order of 0.1 $\mu m^2$ [62] in area, which is a factor 600x smaller than our PairSTDP, 1059x smaller than TripletSTDP, 5383x smaller than RSTDP, and 12699x smaller than a BCPNN synapse. Granted, our silicon synapses run multiple times faster (opening up the possibility to time-multiplex synapse data paths, which is what many neuromorphic designs do) at the expense of reconfiguration. In short, creating capable, capacious, and plastic synapses that bridge the gap between biological systems and silicon is more of a challenge than placing a lot of neurons on a chip.

[5] Where, to be fair, synapses are 3D and not 2D as the silicon we produce, so the comparison is not entirely faithful.

### 3.3 Fastest Level-A (LIF-based) *C. elegans* Simulation

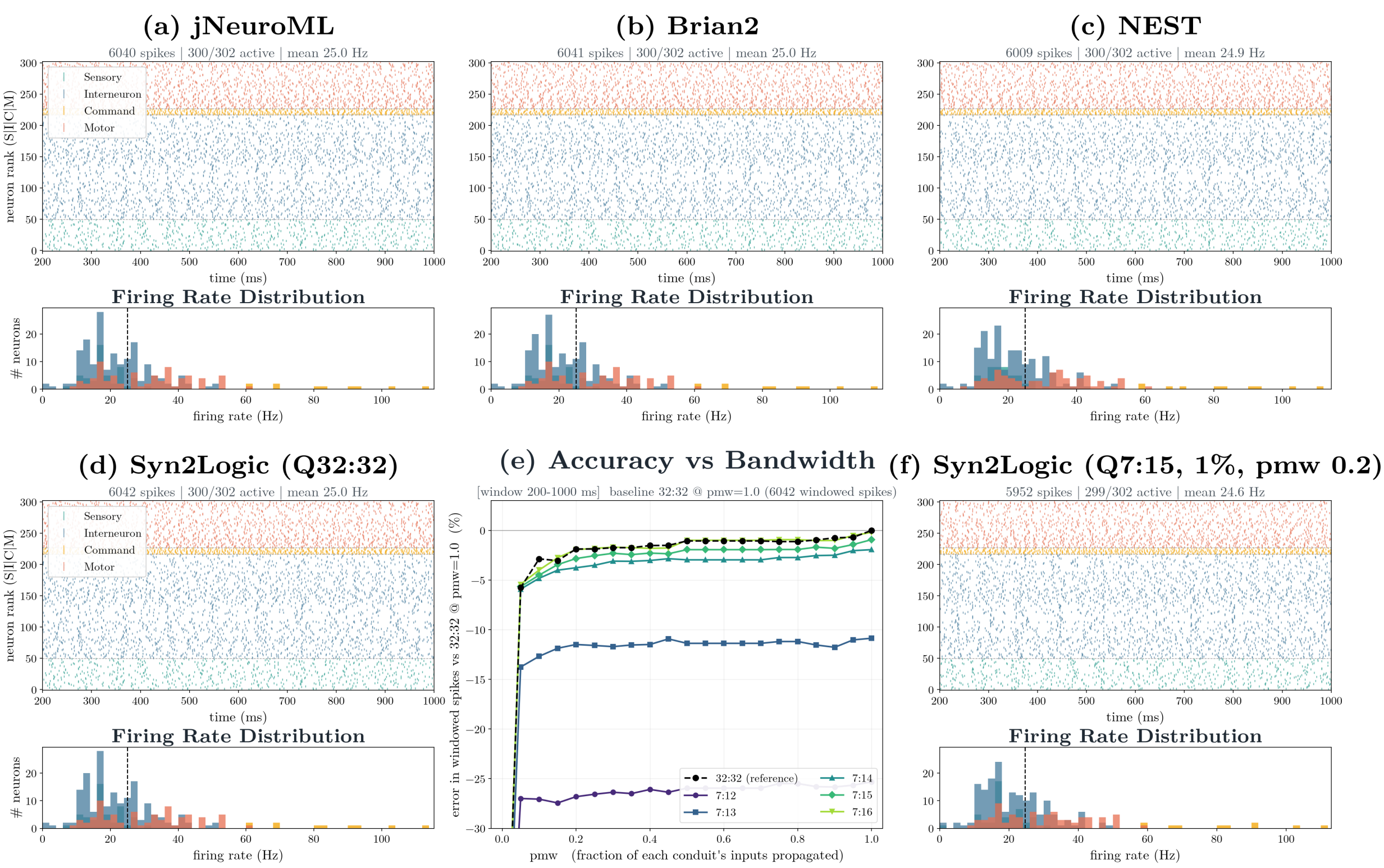

**Fig. 6**: (a-c) Reference jNeuroML, Brian2, and NEST, spike plots for the *C. elegans* simulation, (d) the $Q_{32:32}$ full-precision Syn2Logic spike plot, (e) the impact on data-path width and per-model bandwidth on the error rate, and (f) the final hardware-friendly neuromorphic system.

Next, we turn our attention to the full-scale *C. elegans* connectome shown in Figure 6. Figure 6:a-c show spike raster plots for the network running 0.8 seconds of simulated time [6] on the well-established simulators jNeuroML, Brian2, and NEST, respectively. The different simulators produce near-identical firing rates (25 Hz for NeuroML and Brian2, 24.9 Hz for NEST) but with a different firing rate distribution and a (more or less) equal number of total spikes (6040, 6041, and 6009 for jNeuroML, Brian2, and NEST, respectively). Our $Q_{32:32}$ Syn2Logic description – in Figure 6:d – of the *C. elegans* matches the results of jNeuroML and Brian2, showing 25 Hz mean frequency, 6042 spikes, and a near-identical firing rate distribution. Unfortunately, such a large $Q_{32:32}$ variant would not fit on the FPGAs we use, and so there is a need to design-explore to reduce the size of the final accelerators.

We explored parameters that govern the fixed-precision size and the per-model bandwidth, with the first being motivated by our earlier neuron sensitivity study (Section 3.1) while the second being motivated by the fact that biological SNNs – such as the present *C. elegans* – are very sparsely active, where each neuron (on average) is active only 0.25% of the timesteps– a clear sign that each neuron does not need full bandwidth from each of its synapses. Figure 6:e shows the trade-offs, showing the error (in % of spikes) as a function of per-model bandwidth for a number of fixed-point variants. We see that already at 90% bandwidth reduction (PMW=0.1), most variants reach <5% of error compared to the $Q_{32:32}$ reference, where using 80% bandwidth manages to reach <2.5% error for some configurations. $Q_{7:12}$ and $Q_{7:13}$ variants have too few fractional bits to perform well, irrespective of per-model bandwidth, and instead top out at 25% and 12% error, respectively. A $Q_{7:15}$ variant with 1% constant rounding strikes a strong <1.5% error rate, keeping only 20% of its incoming connections, whose spike raster plot we see in Figure 6:f with 5952 spikes and a 24.6 Hz mean neuron firing rate. As for rounding constants, anything larger than 1% rounding significantly impacts the results negatively.

[6] The full simulation is 1 second, but the first 200 ms the network is mostly silent before activity starts.

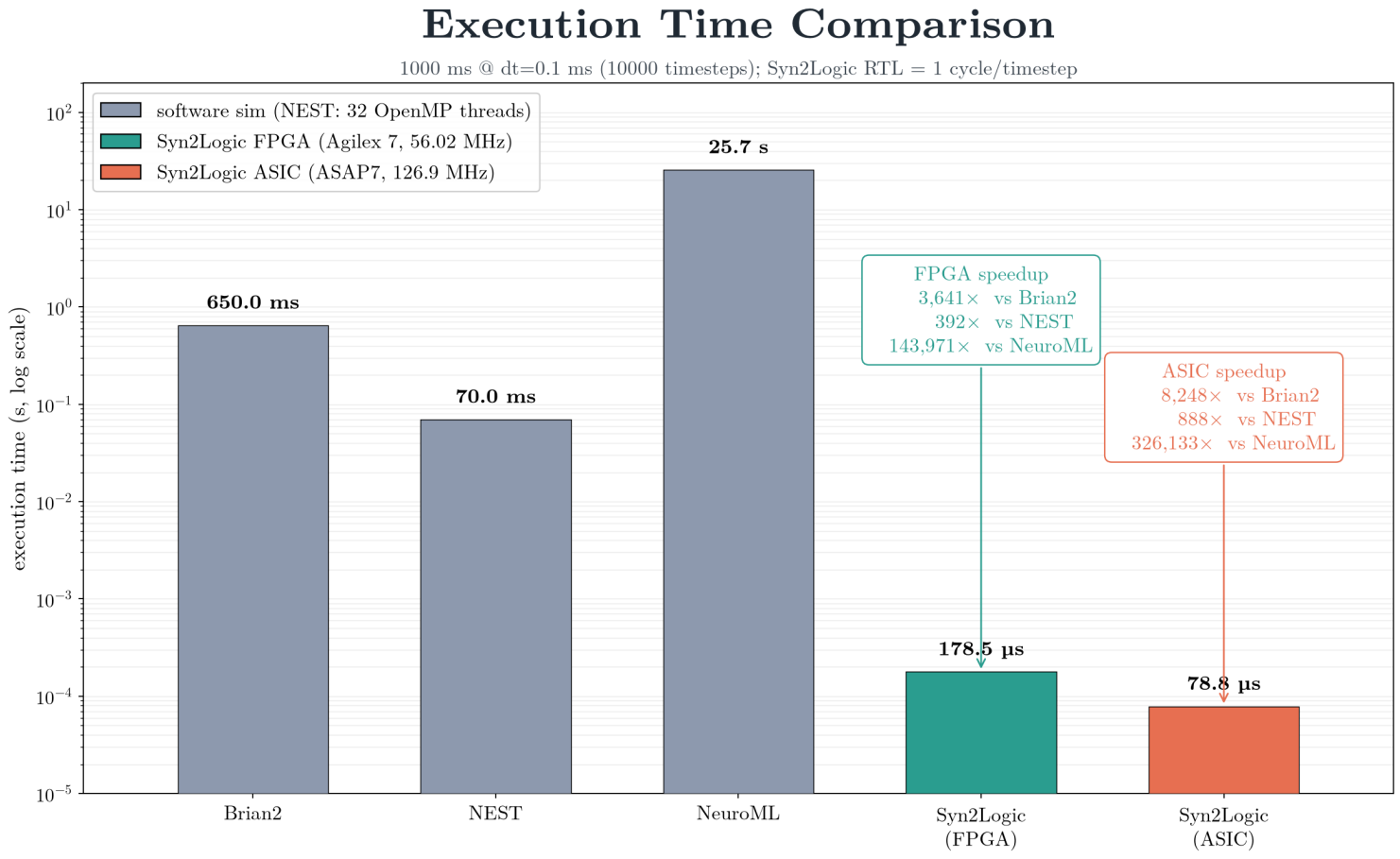

**Fig. 7**: Performance comparison between simulation and Syn2Logic FPGA and ASIC auto-generated *C. elegans* accelerators.

**Table 3**: FPGA and ASIC Synthesis information for *C. elegans*

| **Agilex7** | |
|---|---|
| Logic utilisation (ALMs) | 219,330 / 487,200 (45 %) |
| Dedicated logic registers | 55,420 |
| DSP blocks | 3,926 / 4,510 (87 %) |
| Clock frequency | 56.02 MHz |
| **ASAP7** | |
| Std-cell area | 277,499 $\mu$m$^2$ (0.277 mm$^2$) |
| Cell count | 2,808,514 |
| $f_{\max}$ | 126.9 MHz |
| Total power | 56.4 mW |

Table 3 shows synthesis results for the $Q_{7:15}$ version with 20% per-model bandwidth and 1% rounding, which fits on our Agilex7 FPGA device, occupying half the logical resources and most (87%) of the DSP resources, making it DSP-bound. It reaches a peak clock frequency of 56.02 MHz. The same design, synthesized using Cadence Genus for the ASAP 7nm PDK, yields a 0.277 $mm^2$ design (280,8514 standard cells) with a peak clock frequency of 126.9 MHz.

In Figure 7, we put our accelerator performance numbers into perspective against current state-of-the-art simulators; we see that our FPGA accelerator can simulate the *C. elegans* connectome 392x, 3,641x, and 143,971x faster than NEST, Brian2, and jNeuroML, respectively, running on a large multithreaded server. Furthermore, should our ASIC *C. elegans* accelerator ever be taped out, we would observe a 2.26x speed-up over the FPGA results. To the best of our knowledge, we have shown how to use Syn2Logic to obtain the fastest level-A (LIF-based) *C. elegans* simulation existing today. Our FPGA accelerator runs about 5,600x (ASIC: 12690x) faster than real-time– performance levels that could eventually open up research into understanding plasticity in the worm, neuromorphic and neuroAI crossover, or (in the future) closed-loop experiments.

#### 3.3.1 Discussion: eNDA for use in Neuroscientific Simulation

The speed of neuroscientific simulation is considered one (of four) grand challenges [38] in the pursuit of a deeper understanding of how the brain functions. In particular, faster-than-real-time simulations allow researchers to study the long, slow processes that govern many complex dynamics in the brain. Despite its simplicity, the *C. elegans* model is still not yet fully understood, and by using our **eNDA**-flow to produce high-performance *C. elegans* accelerators that execute between 392x and 143,971x faster than simulators, we can facilitate faster-than-real-time exploration of *C. elegans* , allowing plasticity experiments, neuromorphic/neuroAI crossover, or even (in the future) closed-loop experiments. In the future, we want to extend our **eNDA**-backend to support larger connectomes, and to accelerate the recently charted connectome of the *Drosophila melanogaster* [56] (the common fruit fly).

### 3.4 A Fast Neuromorphic Sudoku Solver

Now we turn our attention towards using SNNs to solve practical optimization problems. One such problem is Sudoku, which represents a canonical constraint satisfaction problem (CSP), with a structure of 81 variables, each containing a digit between 1 and 9, and constraints that enforce all-different on rows, columns, and 3x3 quadrants. The Sudoku problem is NP-complete and has a form that can be adapted to other problems (scheduling, resource allocation, etc.). Furthermore, it is a standard benchmark to solve. Finally, while prior work has shown SNNs to be capable of solving Sudoku, so far *none has ever been able to compete with existing ILP/CP solvers or custom solutions*.

SNN Sudoku Solver

Design and Evaluation on 46 Puzzles

(a) Sudoku Solver Block Diagram

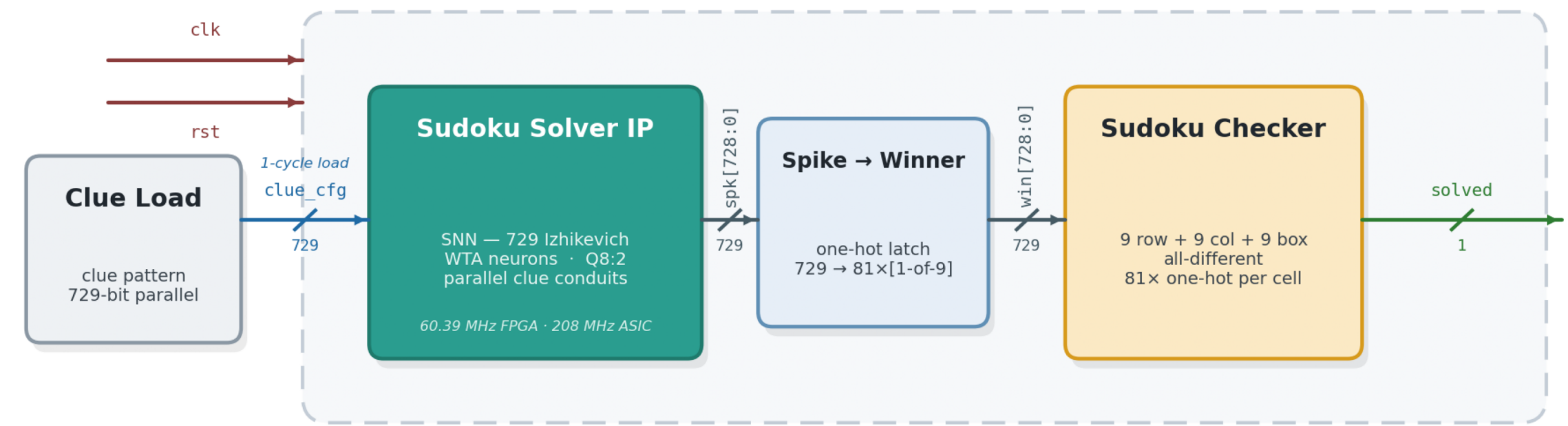

(b) Performance Compared to SOTA

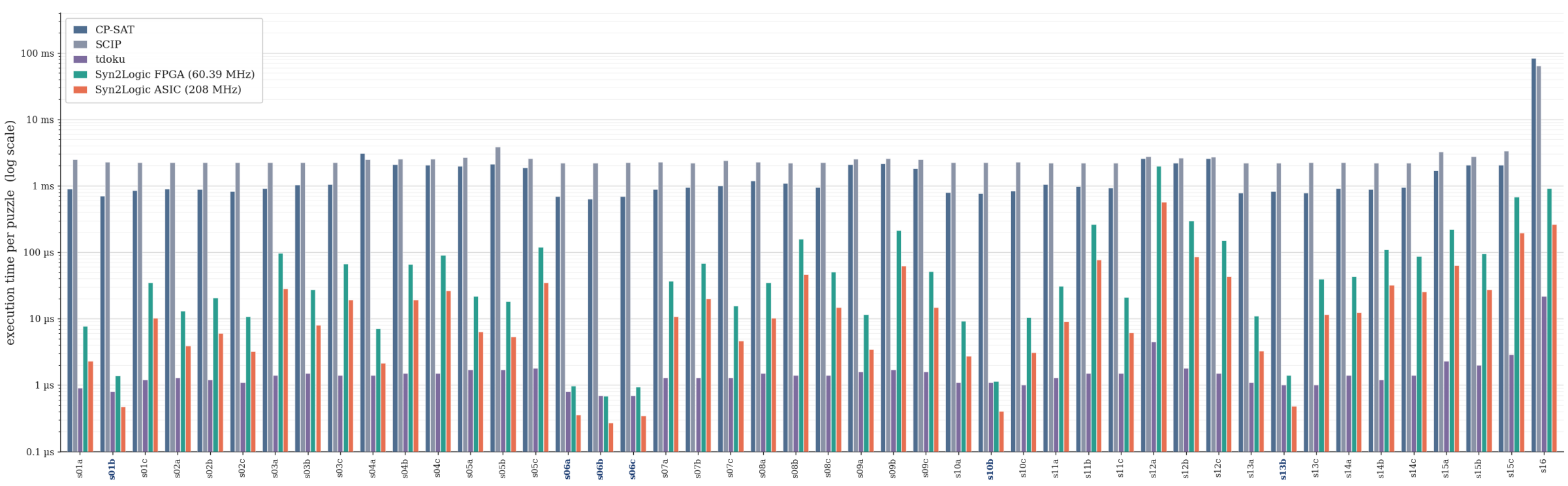

**Fig. 8**: (a) The hardware architecture for solving Sudoku puzzles using our Syn2Logic accelerator, and (b) the performance comparison between CP-SAT, SCIP, tdoku, and our Syn2Logic accelerator per-puzzle solution time.

To solve Sudoku, we designed a network based on the Izhikevich model with added local long- and short-term annealing (escape currents) to avoid being stuck in local (incorrect) minima—a well-known problem in SNN-based solvers. In addition, we performed heavy parameter search to find fast-performing parameters. In the end, the parameters we found working well were: $a = 0.0267, b = 0.25, d = 7.06$, $\Delta t = 2.382$, $I_{\text{const}} = 14.82$, $I_{\text{set}} = 20.77$, $w_{\text{syn}} = 13.5$, $\tau_s = 12.58$, $\Delta\theta = 0.398$, $\tau_\theta = 90.5$, $R_{\text{amp}} = 17.63$, $\tau_r = 65$, $v_{\text{th}} = 32$, $c = -65$, which were then quantized and optimized for a $Q_{8:2}$ format.

There are a few very interesting things with the above parameters. Firstly, we treat the integration time step as a free parameter to optimize, which resulted in a very large time step of 2.382 ms — more than 20x larger than traditional neuroscientific simulations of the Izhikevich model ($0.1ms$). Secondly — and this is counterintuitive — we found that the model worked best with a very small number of fractional bits, and actually a $Q_{8:2}$ format was shown to perform better than higher-precision alternatives. These two observations run counter to standard practices for using SNNs to solve optimization problems. Furthermore, these optimizations — in particular the second one — are very hardware-friendly.

**Table 4**: FPGA and ASIC synthesis (ASAP7 7 nm, Genus synthesis estimations) results for our Sudoku accelerator.

| **Agilex 7** | |
|---|---|
| Logic utilization (ALMs) | 426,278 (87%) |
| Dedicated logic registers | 75,315 |
| DSP blocks | 1,823 / 4,510 (40%) |
| Block RAM / memory | 0 (none used) |
| Clock frequency ($F_{\text{max}}$) | 60.39 MHz |
| **ASAP7** | |
| Standard cells | 2,033,761 |
| Cell area | 177,086 $\mu$m$^2$ (0.177 mm$^2$) |
| Total power | 61.3 mW |
| Clock frequency ($F_{\text{max}}$) | 208.2 MHz |

The accelerator we designed is shown in Figure 8:a, and consists of a WTA circuit for solving and enforcing the Sudoku constraints, configurable externally by sending spikes to the conduit (to set Sudoku clues), and updates the network one timestep per clock cycle. Each timestep, all 729 neuron spike outputs (if any) are latched into the checker network, which rejects invalid puzzles. In short, one component is searching for plausible solutions, and the other component rejects invalid solutions. The spike checker networks are negligible in area and are not on the critical path.

Synthesis of our accelerator is shown in Table 4. On the FPGA, the accelerator is logic-bound, consuming 87% of device resources and 40% of all DSP resources, running at a peak clock frequency of 60.39 MHz. A hypothetical ASIC variant using the ASAP 7nm PDK and Cadence Genus estimations would consume order 2 million standard cells, be 177k $\mu m^2$, and run at 208.2 MHz.

Our accelerator performance, compared to the state-of-the-art optimization solvers CP-SAT, SCIP, and the handwritten `tdoku` solver (claimed to be the world's fastest solver), is shown in Figure 8:b on all 46 Mantere–Koljonen puzzles [63]. The Mantere–Koljonen puzzles are a mix of easy to hard puzzles, and include the notoriously hard AI escargot puzzle (s16).

Results show that SCIP [30], CP-SAT [31], and tdoku [32] solve all 46 puzzles correctly in 173.2 ms, 142.5 ms, and 82.3 $\mu$s, respectively, on a modern server machine. Our Syn2Logic FPGA accelerator solves all 46 puzzles correctly within 6.19 ms, and the hypothetical ASIC variant within 1.80 ms– many times faster than both SCIP and CP-SAT. While tdoku is still much faster than Syn2Logic, our accelerator solves some individual puzzles faster than tdoku, such as s01b, s06a, s06b, s06c, s10b, and s13b. It is not unlikely that for these puzzles, our accelerator is the fastest solver currently in the world.

### 3.4.1 Generalization to other puzzles

Does our neuromorphic solver generalize outside the very same puzzles we tuned it for? Does it still perform well? To answer this, we first tested it on a different but equally hard Sudoku testing set, the Euler-96. Our solver solves 42 (the Euler-96 has 50 puzzles, but we removed overlaps with Mantere–Koljonen) of these puzzles, consuming 2.79 ms/ 0.81 ms (FPGA/ASIC) for the whole set and between 0.51 $\mu s$/0.17 $\mu s$ and 427.8 $\mu s$/ 124.1 $\mu s$ per-puzzle solution time. CP-SAT and SCIP solve the entire set in 49.5 ± 22.0 ms and 96.7 ± 4.3 ms, respectively, and our solver continues to be 17.74x/61x (FPGA/ASIC) and 34.65x/119x (FPGA/ASIC) faster than CP-SAT and SCIP, respectively. tdoku continues to be the fastest with a total time of 0.073 ms, which is 11x faster than our ASIC solver; our ASIC solver solves five puzzles (out of 42) 1.16x-4x faster than tdoku.

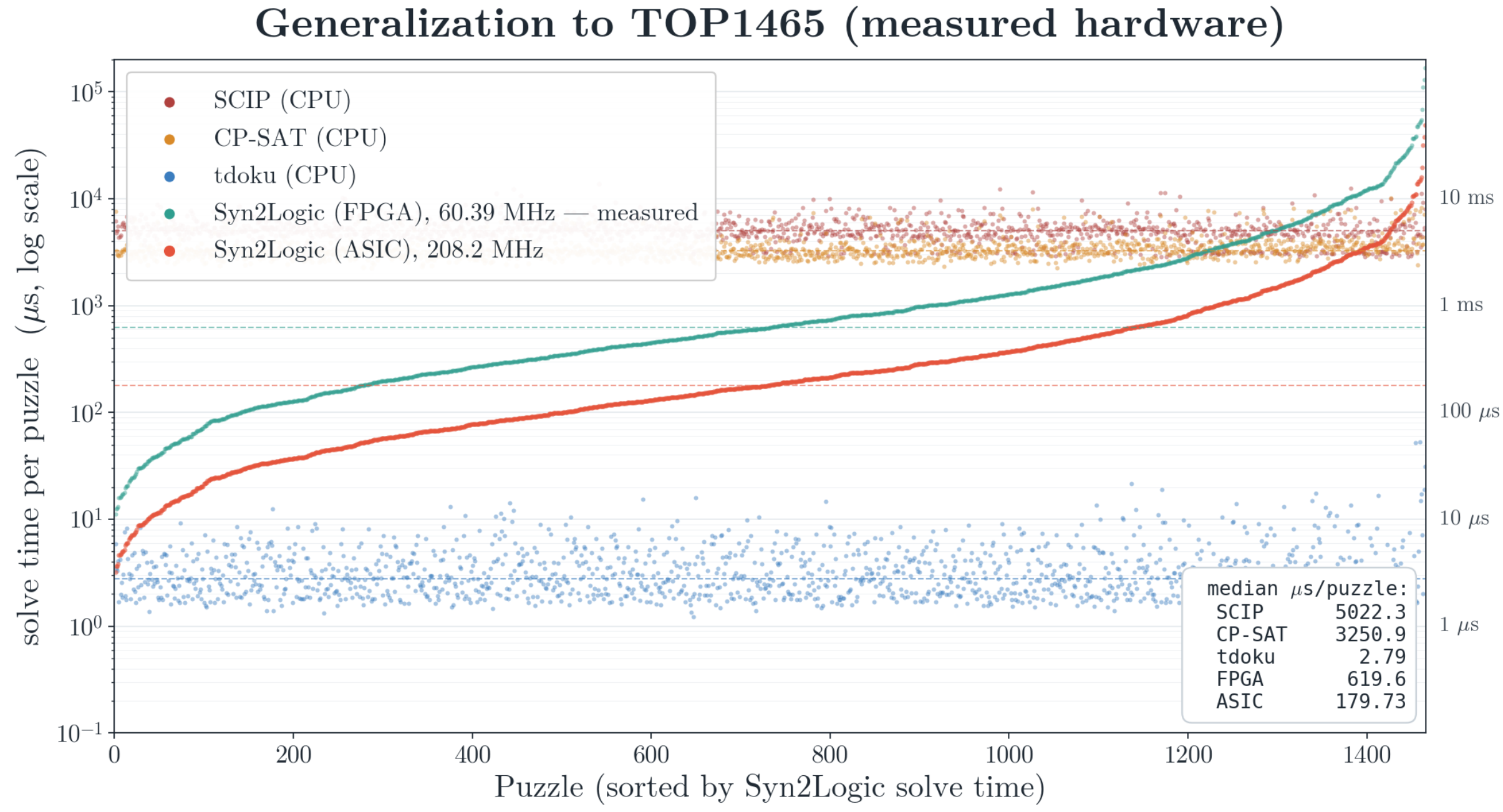

**Fig. 9**: Generalization testing of our Sudoku solver on puzzles and in comparison with the software solvers.

The ultimate test of generality is whether our solver can solve the top1465— a set of 1465 handpicked hard puzzles [64]; these puzzles are significantly harder ( 5-35x based on the number of guesses needed) than our previous tests. Results are shown in Figure 9. Our solver solves **all** 1465 puzzles in 3.90s/1.13s (FPGA/ASIC), dominated by a set of 15 puzzles that need>2 million steps to solve, with the main culprit (puzzle p0330) taking 10 million cycles to solve, and roughly half (43.8%) the total time is spent on 50 (out of 1465) hardest puzzles. CP-SAT and SCIP both solve the entire puzzle set in 5095.8 ± 6.2 ms and 7516.1 ± 153.1 ms, respectively, reducing the speed-up our solver has over CP-SAT to 1.31x/4.5x (FPGA/ASIC) and SCIP to 1.93x/6.64x (FPGA/ASIC)— still a large improvement, but reduced compared to the simpler puzzle. tdoku continues to be blazing fast, with a total time solving all puzzles of 5.3 ± 0.004

ms– two orders of magnitude faster than our solver variants. We only solve four puzzles (out of the 1465) between 1.07-1.83x faster than tdoku.

### 3.4.2 Discussion: eNDA Enables High-Performance Practical SNN-based Optimizers

Until now, prior neuromorphic accelerators and simulators had a per-puzzle solution time in order of seconds [65]; our work shows that we can employ **eNDA** to create neuromorphic accelerators that have as low as sub-$us$ puzzle solver time and outperform state-of-the-art optimization solvers (CP-SAT/SCIP) by up to 28x on an FPGA and 96.2x if an ASIC were manufactured. We also show – for the first time and for a *small* subset of puzzles – that a neuromorphic solver *can* outperform the world's fastest Sudoku solver (tdoku) running on a state-of-the-art server-class system on some puzzles. We also showed that our solver generalizes, and solves harder puzzles than what it was tuned for, solving the entire top1465 puzzle list. Aside from the **eNDA**-flow, which is central to the success of the hardware accelerator, there are two very interesting observations: **(i)** our neuron model actually worked the best with a very large integration time step of over $2ms$, and **(ii)** a small fixed-point representation actually allowed (for this particular configuration) to solve more puzzles than its larger counterpart. Our work paves the way toward enabling neuromorphic optimization solvers to be a serious alternative to traditional solvers, and to be far more energy-efficient (ASIC estimates are $61.3mW$, which is 3-4 orders of magnitude lower than the server-class processor we compared against), making them suitable for serious exploration of real use-cases such as Edge-User Allocation [66] or scheduling sports events and/or airplanes.

A final note on the configuration of the puzzle: our experiments did not take into account the binary loading overhead of the software solver. Our accelerator can load a puzzle instance in a single cycle, so likely – if we also include the time to reconfigure a new puzzle – the results would favor our accelerator even more.

## 3.5 Energy-Efficient MNIST Accelerator

We continue our results with a familiar use case: identifying handwritten digits using the MNIST dataset. While not the largest, the MNIST dataset still represents a cornerstone for evaluating — and more importantly, comparing — neuromorphic systems with each other. The dataset comprises 60000 training and 10000 test images of 28x28 pixels that are black and white, where each corresponds to the written value 0-9. Our goal was to create the most energy-efficient neuromorphic MNIST classifier to date, runnable on hardware that is readily available and cheap (in contrast to the Agilex7 board in prior sections). We achieved this through our neuromorphic design automation to make an MNIST accelerator for a system *that is financially accessible to a broad audience*, such as the MAX10 toy-FPGA, released in **2014** and built using a **55 nm** technology.

After training a 784-128-64-10 layer MLP network containing 110k parameters using snnTorch, we start our design-space exploration. The initial network demonstrated 97.61% testing accuracy, well in line with current state-of-the-art. Next, we will explore the design space of the network.

We start with the impact of sparsity of the network, in Figure 10:a. We noticed that the network can be pruned by 80-90% of existing parameters with minimal impact on classification accuracy, especially when coupled with fine-tuning of the network after pruning, yielding an accuracy of 97.23% and 95.06% test accuracy for 80% and 90% pruned networks, respectively. While this is a respectable amount of pruning, in fact, such a network is still far too large to fit on the 50k look-up tables on the MAX10 board, due to the many adders that make up reduction trees per neuron.

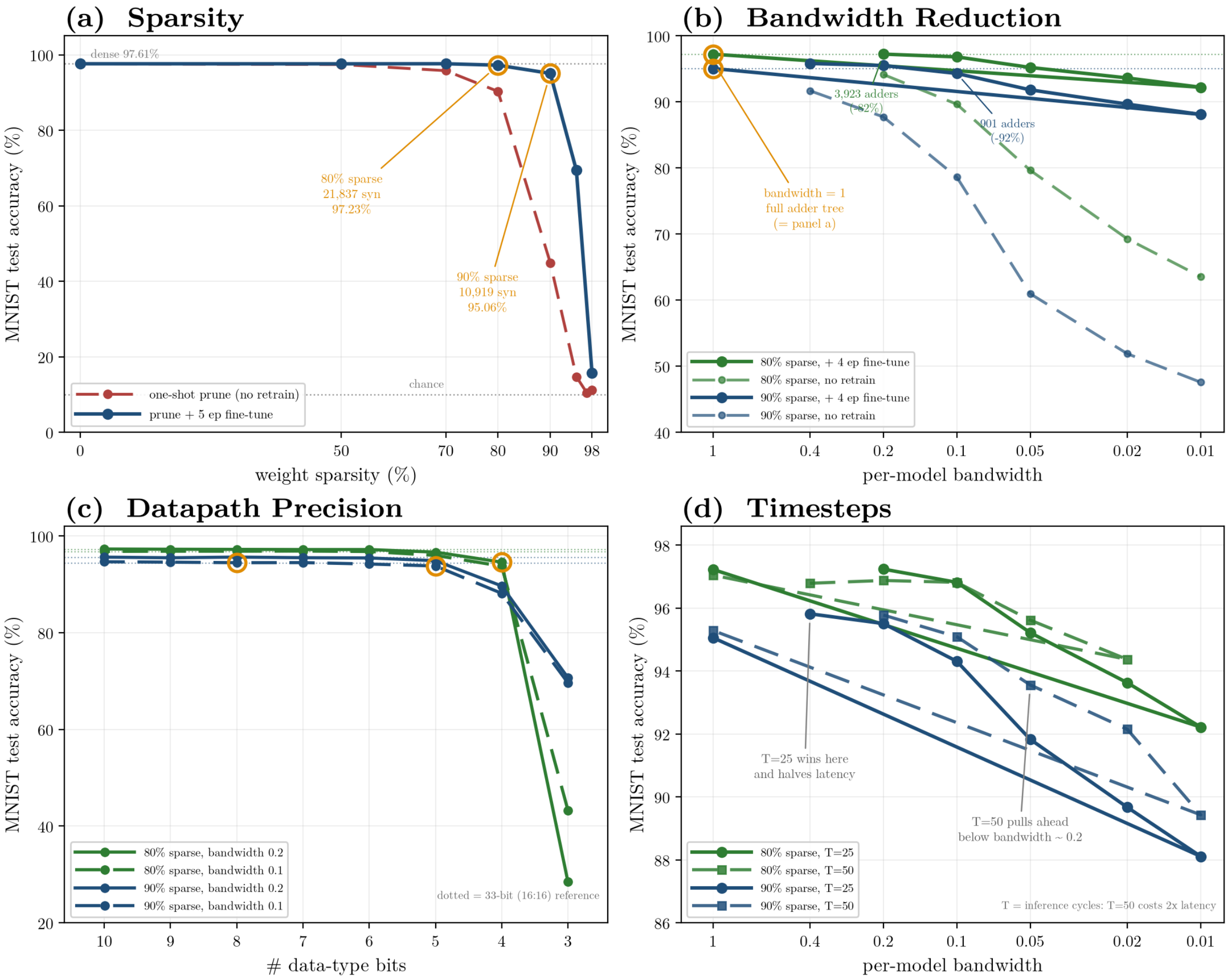

**Fig. 10**: (a) Impact of weight-pruning on our network, (b) impact of our per-model bandwidth parameter on the pruned network, (c) impact of reducing the SNN data-path width on the network, and (d) impact of timesteps on the accuracy and per-model bandwidth.

Next, we investigated the impact of per-model bandwidth reduction on the already sparse network, in Figure 10:b. We noticed that we can reduce the per-model bandwidth between 80% and 90% without losing much test accuracy, again exploiting fine-tuning to reach between 97.24% (upper, interestingly marginally better test accuracy after fine-tuning than pmw=1.0/80% baseline) and 94.31% (lower) test accuracy. By reducing the per-model bandwidth, we effectively reduce the number of adders required in the reduction step per neuron by  80% (pmw=0.2) and  90% (pmw=0.1).

Having sparsified and reduced the per-model bandwidth of the network, we next investigate the impact of data-path size on our network, in Figure 10:c, by plotting and investigating the test accuracy as a function of data-type size. We noticed that we can reduce the size of operations to as small as $Q_{1:4}$ bits with small impacts on the accuracy (97.20%), compared to our $Q_{16:16}$ reference. This means that we can still achieve 93%+ accuracy with a $Q_{1:3}$ format. Interestingly, $Q_{1:8}$ reaches 97.31%.

An interesting observation and experiment can be seen in Figure 10:d, where we investigate the impact of per-model bandwidth as a function of how many timesteps we run the network and input for. An interesting observation is that, as the timestep increases from 25 $\rightarrow$ 50 timesteps, it becomes more resilient to degradations from a reduced bandwidth per model. This is intuitive, since longer timesteps can lead to sparser activations in the network, which in turn lead to fewer spikes incoming to each neuron. The interesting part is that, by controlling the sparsity of the network, this impacts hardware generation significantly, since at a 90+% reduction in per-model bandwidth together with the earlier sparsity pruning, most of the adder trees are gone, and replaced with only the priority encoders and multiplexers.

**Table 5**: FPGA and ASIC synthesis results for the MNIST accelerator.

| | **MAX10** | **Agilex7** | **ASAP7** |
|---|---|---|---|
| **Design point** | | | |
| Weight sparsity / synapses | 90 % / 10,919 | 80 % / 21,837 | 80 % / 21,837 |
| Per-model bandwidth (`--pmw`) | 0.1 | 0.2 | 0.2 |
| Datapath (`--fixed-point`) | $Q_{1:3}$ (5 bits) | $Q_{1:4}$ (6 bits) | $Q_{1:4}$ (6 bits) |
| Timesteps $T$ (= cycles/image) | 25 | 25 | 25 |
| Accuracy, full 10k test set | 93.80 % | 97.20 % | 97.20 % |
| **Implementation** | | | |
| Logic | 38,103 LE (77 %) | 26,684 ALM (5 %) | 110,146 std cells |
| of which the classifier | 18,759 LE | all | all |
| Registers | 12,498 (classifier 1,074) | 4,077 | — |
| On-chip memory | 33,744 bits (92 M9K) | — | — |
| Cell area | — | — | 8,765 $\mu$m$^2$ |
| die | — | — | 21,764 $\mu$m$^2$ |
| **Closure and cost** | | | |
| Clock constraint | 43 MHz | 400 MHz | 500 MHz |
| $F_{\max}$ | 45.28 MHz | 214.82 MHz | 491 MHz |
| Power | 307 mW | — | 17.96 mW |
| Latency per image | 581 ns | 116 ns | 51 ns |
| Energy per image | 179 nJ | — | 0.91 nJ |

(a) **Floorplan** (b) **In-Lab testing**

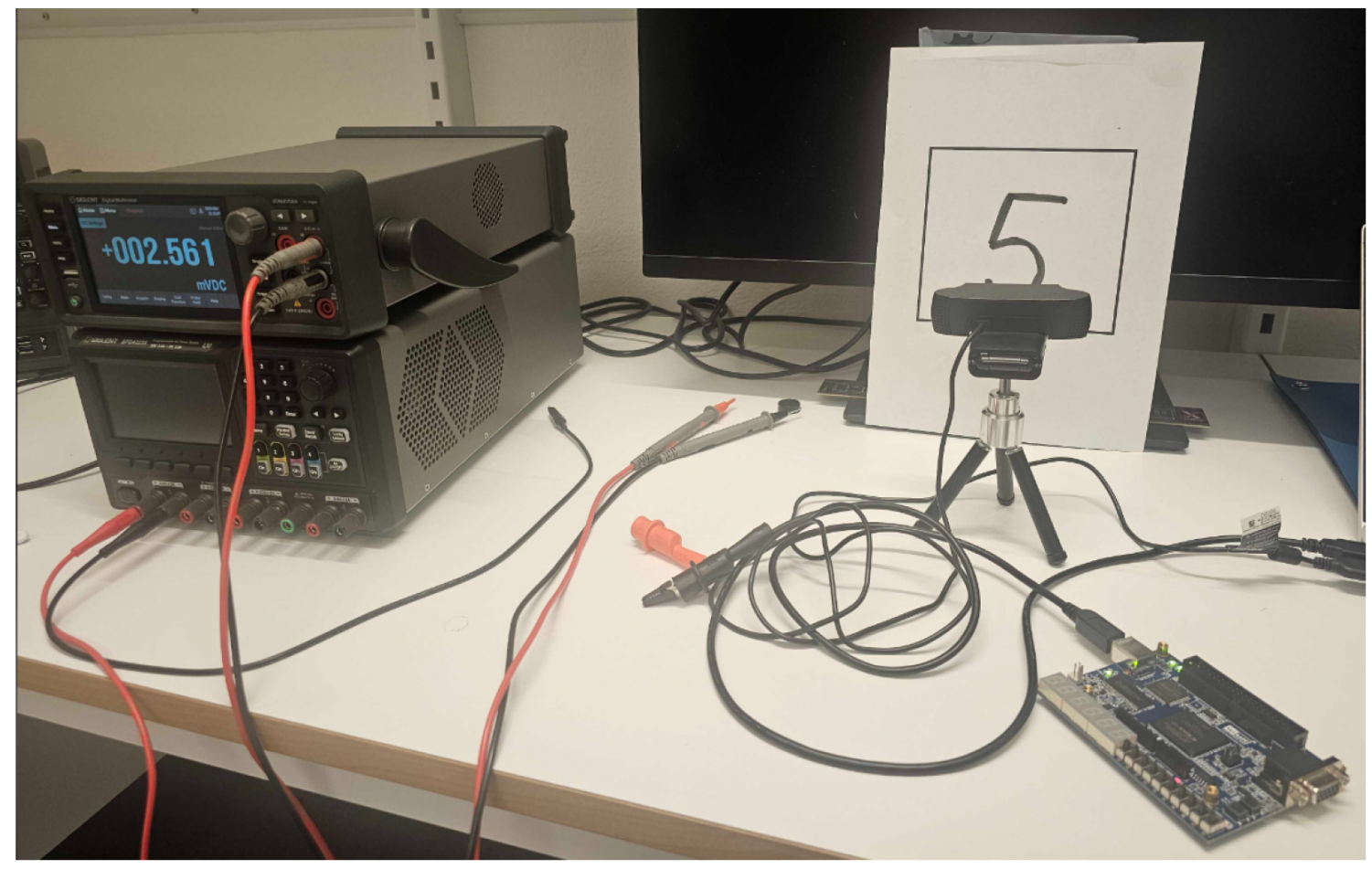

**Fig. 11**: (a) the ASIC floorplan for the placed-and-routed chip (each neuron is clearly shown), and (b) just after the power measurement for the MAX10 device when inferring a hand-drawn number 5 (the observant reader will see the fifth LED shining strongest, indicating a correct inference).

At this point, we can start synthesizing our accelerators, with results shown in Table 5. We created three different variants, suitable for MAX10, Agilex7, and ASIC ASAP7 generation. The MAX10 variant uses the 90% sparse network in $Q_{1:3}$ format with 90% (T=25) per-model bandwidth reduction and reaches 93.80% testing accuracy, consuming 77% of device resources and no DSPs. The MAX10 power consumption was empirically measured to be 0.307 Watt. For the larger Agilex7 system, we show results for a variant that uses the 80% sparse network with $Q_{1:4}$ format with 80% per-model bandwidth reduction and reaches 97.20% test accuracy, consuming a minimal fraction of the resources (5%). Finally, we show synthesis results for the ASAP7 accelerator that uses the same configuration as the Agilex7 variant, consumes 110k standard cells, and is estimated at 17.96$mW$ power consumption. The ASIC floorplan is shown in Figure 11:a. Performance-wise, the ASIC version is the fastest at 491 MHz, followed by the Agilex7 at 214 MHz and lastly the MAX10 at 45.28 MHz. An image of the MAX10 power-consumption measurement evaluation (or directly after, actually, with the snapshotted measurement still visible) inferring a hand-written number 5 (LED #5 is lit) is seen in Figure 11:b.

* Our calculations based on publication data

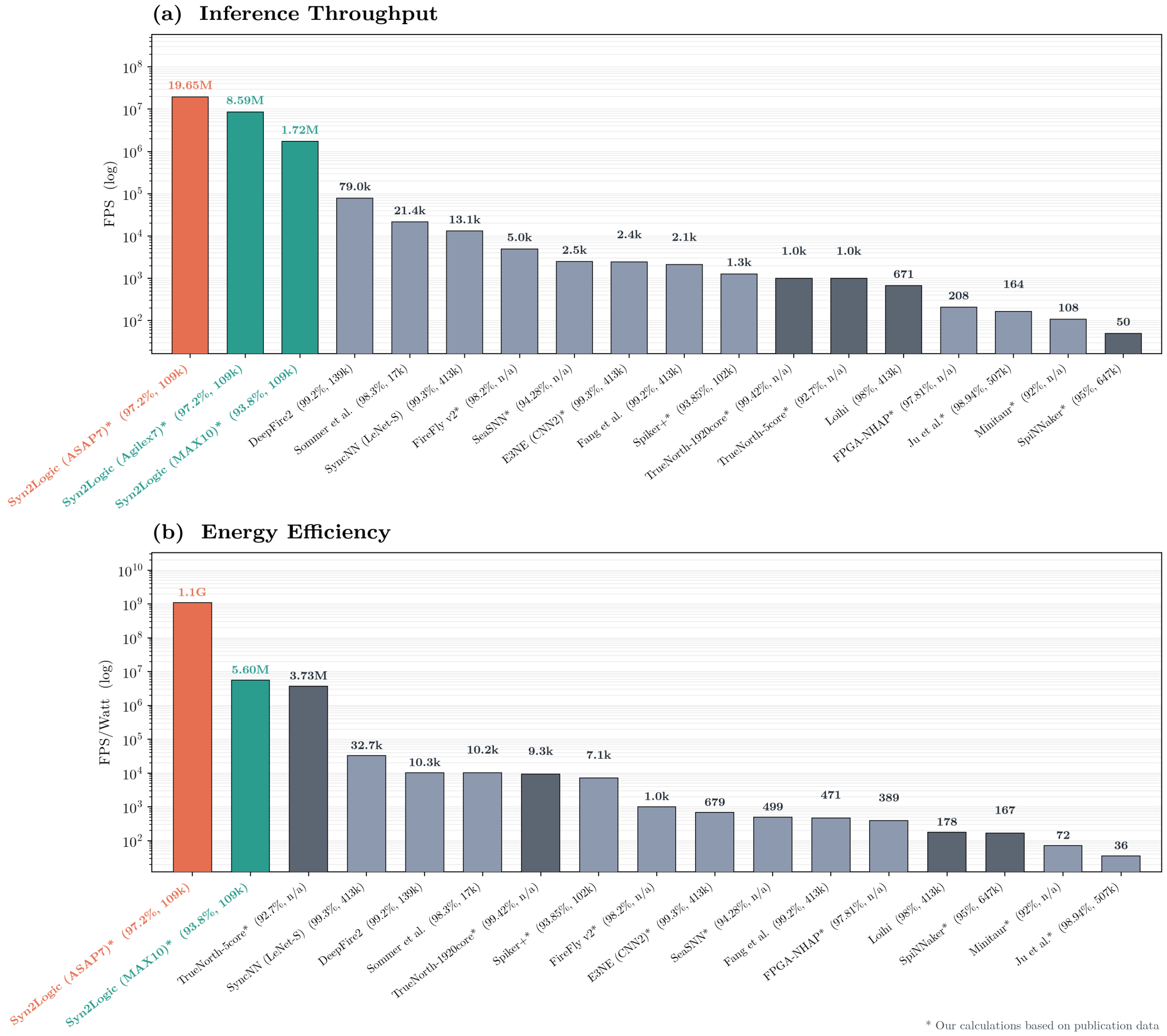

**Fig. 12**: (a) Frame-per-second (FPS) performance of our Syn2Logic accelerator over state-of-the-art neuromorphic systems, and (b) FPS/Watt over state-of-the-art systems.

The performance (FPS) of the accelerators and their energy efficiency (FPS/Watt) are shown in Figure 12:a-b and are compared with the reported performance of state-of-the-art SNN accelerators inferring MNIST. Performance-wise, our Agilex7 and ASIC accelerators are up to 109 and 249x faster than DeepFire2, respectively, while the MAX10 version is 22x faster than the runner-up DeepFire2. From an energy-consumption perspective, the ASAP7 performance could be up to 1.1 GFPS/Watt, while the MAX10 measured energy consumption is 5.60 MFPS/Watt— 1.5x more power-efficient than the IBM TrueNorth runner-up. Remaining SNN accelerators are 32.7 kFPS/Watt and below.

In Figure 13 we see a different view of the state-of-the-art landscape, where we plot power-efficiency as a function of inference classification accuracy (%). We see that, for the inference accuracy Syn2Logic accelerator provides, they are significantly (many orders of magnitude) more energy-efficient than alternatives, even when compared to other accelerators that provide the same (or less) inference accuracy. While there are several accelerators reaching a higher accuracy than our 97.20%, those 1-2% come at a four-order-of-magnitude cost in battery drainage. We have to acknowledge that the ASAP7-variant is an estimate for the core only. Furthermore, these numbers assume that we can feed the images to the accelerator at the maximal FPS; while this is true for the MAX10-accelerator (that would hypothetically need to be fed with HBM2-quality external memory), there is nothing external that could feed 19.65 million images for the ASIC or 8.59 million images for the high-end FPGA, so these systems are more memory-bound than compute-bound. That being said, there are ample opportunities to tune these accelerators to the data source. For example, if the data source is an event-driven camera operating at 40 FPS, then the clock frequency can be significantly reduced (e.g., 491 MHz → 40 Hz, assuming each frame is given in parallel) for the device to draw significantly less dynamic energy. There is also one dimension not thoroughly explored in our work: the number of timesteps per image. In our work, we used a conservative 25 timesteps per image, while most other work uses between 4 and 8, which linearly reduces the required data bandwidth to provide the data. In short, for a timestep of four, all our performance numbers would be 6.25x higher.

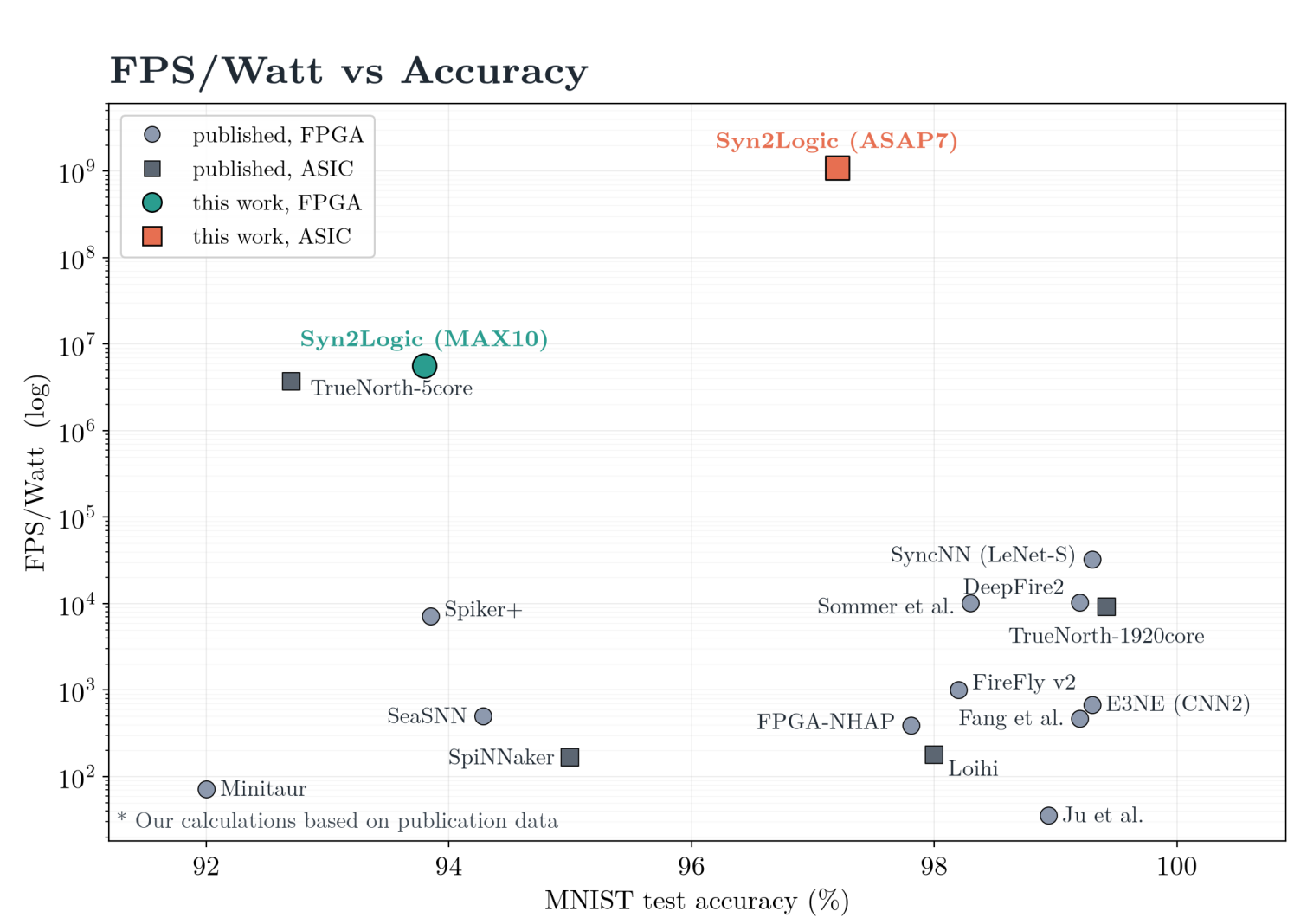

**Fig. 13**: Energy-efficiency versus classification accuracy for different MNIST accelerators.

### 3.5.1 Discussion: eNDA for Extremely Energy-Efficient TinyML on FPGAs

Our accelerator numbers could, to the best of our knowledge, be the most energy-efficient MNIST neuromorphic numbers in the world. Now, this might not be an apples-to-apples comparison, since every one of the reported accelerators used a different network architecture (SCNN, SMLP, etc.) with varying model parameter counts (we used 110k, DeepFire2 139k, Sommer et al. 17k, etc.). With that being said, let the result sink in: we trained an S-MLP network in a matter of hours, explored the design space using Syn2Logic, and created an MNIST classifier that reaches 1.72 Million FPS with an efficiency of 5.6 Million FPS/Watt (full System-on-Chip, not only estimations of the IP module), beating **all current top-end SNN accelerators on MNIST inference**, running **on a toy FPGA** that is significantly cheaper, older (2014), and more constrained than any of the competition *without writing a single line of hardware description language code* using our **eNDA**-flow.

## 3.6 Satellite Telemetry Anomaly Detection

Until now, we have shown how our **eNDA**-flow can be used to push the state of the art on three use cases from different frontiers (neuroscience, optimization solvers, and image recognition). For each of these use cases, the *network* solved the particular problem, which is fantastic from a performance perspective. However, sometimes we do not know the exact network structure that solves a problem or even do not have access to the data itself. Instead, we want something *generic* that can be dynamically trained (on- or offline) and even used for different use cases (and we also want to show the more advanced expressiveness of the Syn2Logic language).

In this use-case, we will create a *generic* neuromorphic component; a component that **(i)** does not encode the problem in the network, **(ii)** has *online* plasticity enabling unsupervised learning, **(iii)** is modular and can be scaled up/down subject to the needed capacity, and **(iv)** is hierarchical in nature.

A picture of the system we have created is shown in Figure 14. At the lowest level, Figure 14:a, we describe a **P**rocessing **E**lement (PE), which consists of a single LIF neuron and a compile-time arbitrary number of STDP synapses. The synapses feed the LIF neuron and receive a post-synaptic spike (PSP) through a feedback path from the neuron. Additionally, each PE has two inhibitory inputs: one semi-local (coming from other PEs) and one global (coming from the global input),

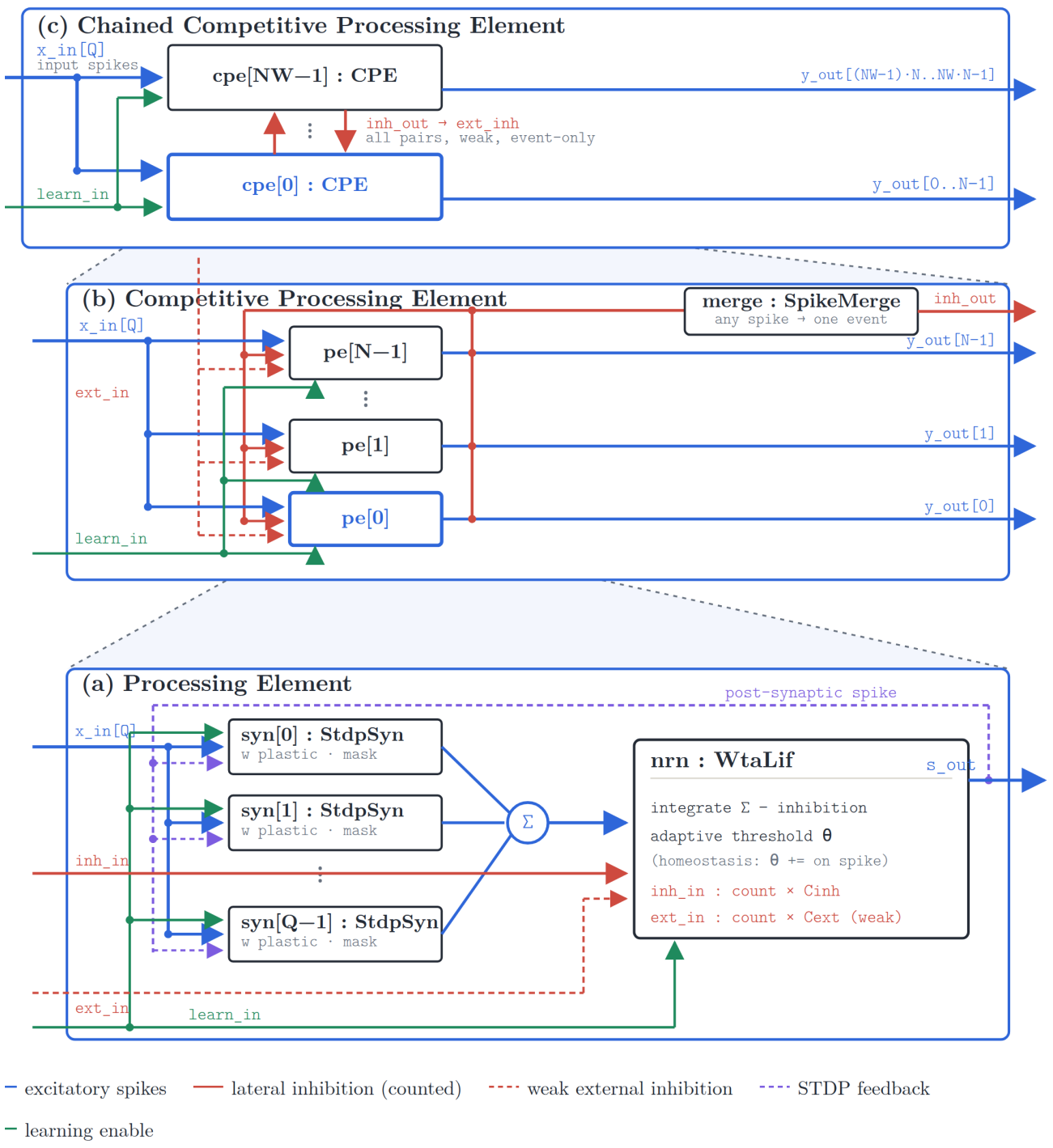

**Fig. 14**: Our generic hierarchical accelerator, showing (a) the bottom-level PE element that is combined in a CPE (b), which can be further chained into multiple CCPE (c).

which can be used to provide competition between
PEs. The next level in the hierarchy, called **C**ompetitive PE (CPE) Figure 14:b, connects multiple PEs together under a single *receptive field* (that is, each PE *reacts* to the same input). Each PE's output spike is inhibitory-connected to each other PE, but also wires out of the CPE component. In addition to the receptive fields (input spikes to the module), we also see the global inhibitory inputs and a global inhibitory signal generated by combining all output spikes from the PEs. A learning signal enables (or disables) learning of the STDP synapses and the threshold of the neuron. Finally, in our final hierarchical layer and Figure 14:c, we **C**hain CPE (CCPE) components to increase the capacity and connect them using the global inhibitory components. Every component in the system can be enabled/disabled, and the entire system in 168 lines of Syn2Logic code.

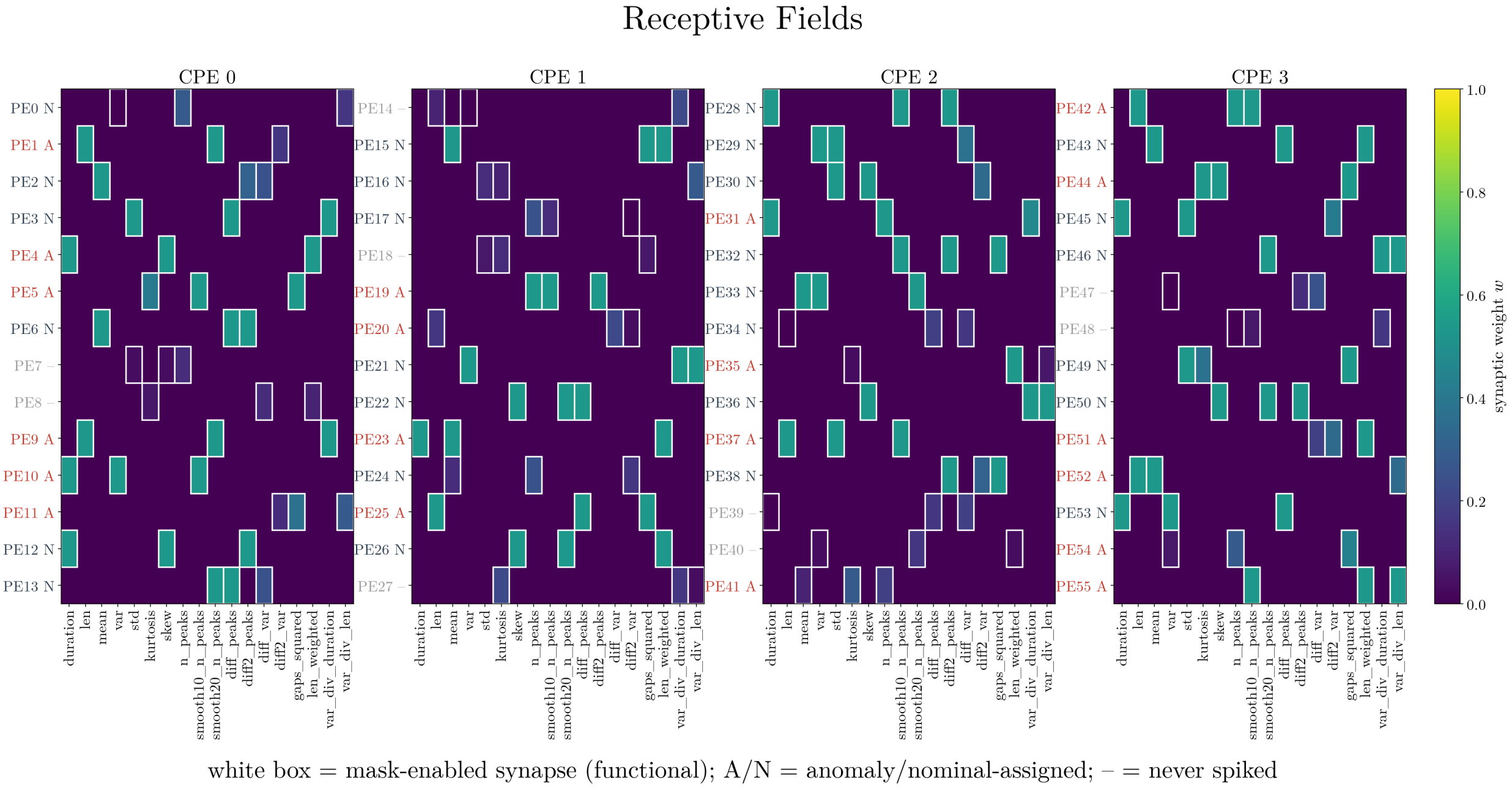

**Fig. 15**: The learned weights for each PE across CPEs and on what telemetry data, showing 20 PEs detecting anomalies, 27 PEs detecting normal activity, and 9 that never spiked during the test.

For demonstration purposes, we created a sample system that (in total) contains 56 neurons, 1008 plastic synapses, and 784 inhibitory synapses inside the CCPE. The CCPE itself has 4 x CPEs, each with 14 PEs, with 18 STDP synapses per PE. We used a $Q_{1:5}$ representation. In total, the system has 1k learnable parameters and 2k tunable parameters. We built an SoC for using our accelerator on the Agilex7 platform; the entire SoC consumed 95% logic resources, 1% of all DSP blocks, and 17% of all on-chip resources (used for storing and saving spike conduits/outputs) [7]. The total configuration scan chain was 38,024 bits, and the system ran at 15 MHz.

We tested the generated system on the OPS-SAT benchmark [67], used to predict whether anomalies have occurred in satellite telemetry data. After training unsupervised for one epoch followed by the benchmark's label-based readout, the system learned fairly well to detect telemetry anomalies.
The internal receptive fields and their weights for each PE is seen in Figure 15 for the four CCPE and 56 PEs, and can see that the system has learned to differentiate the telemetry of the OPS-SAT sensors; the `len` was by far the most important (by summing learned weights on anomaly-assigned neurons) signal to detect anomalies, with runners-up including `smooth10_n_peaks`, `len_weighted`, `gaps_squared` and `duration`. The performance of our accelerator reaches a functional performance of AUCPR=0.712, placing it second in performance in unsupervised performance on this dataset, with MO-GAAL being the best (AUCPR=0.779) and AnoGAN (AUCPR=0.668) [67]. The full comparison (extended with our data from [67]) is seen in Table 6.

**Table 6**: Unsupervised anomaly detection on the OPS-SAT-AD test set, sorted by $AUC_{PR}$ with baselines from Table 3 [67].

| Detector | $AUC_{PR}$ | $AUC_{ROC}$ | Acc | $F_1$ | P | R | MCC |
|---|---|---|---|---|---|---|---|
| MO-GAAL | 0.779 | 0.865 | 0.907 | 0.726 | 0.985 | 0.575 | 0.710 |
| **Syn2Logic** | **0.712** | **0.846** | **0.845** | **0.627** | **0.645** | **0.611** | **0.530** |
| AnoGAN | 0.668 | 0.756 | 0.868 | 0.588 | 0.877 | 0.442 | 0.563 |
| SO-GAAL | 0.660 | 0.749 | 0.885 | 0.655 | 0.906 | 0.513 | 0.627 |
| OCSVM | 0.659 | 0.787 | 0.845 | 0.647 | 0.630 | 0.664 | 0.548 |
| KNN | 0.658 | 0.852 | 0.824 | 0.575 | 0.594 | 0.558 | 0.465 |
| ABOD | 0.644 | 0.843 | 0.832 | 0.582 | 0.620 | 0.549 | 0.479 |
| INNE | 0.643 | 0.806 | 0.847 | 0.646 | 0.638 | 0.655 | 0.549 |
| ALAD | 0.629 | 0.744 | 0.870 | 0.596 | 0.879 | 0.451 | 0.570 |
| LMDD | 0.623 | 0.767 | 0.854 | 0.628 | 0.691 | 0.575 | 0.542 |
| SOD | 0.621 | 0.797 | 0.737 | 0.505 | 0.423 | 0.628 | 0.348 |
| COF | 0.603 | 0.774 | 0.794 | 0.576 | 0.514 | 0.655 | 0.448 |
| LODA | 0.597 | 0.748 | 0.822 | 0.588 | 0.583 | 0.593 | 0.475 |
| LUNAR | 0.540 | 0.792 | 0.813 | 0.407 | 0.630 | 0.301 | 0.342 |
| CBLOF | 0.493 | 0.642 | 0.756 | 0.427 | 0.429 | 0.425 | 0.272 |
| DIF | 0.465 | 0.797 | 0.790 | 0.035 | 1.000 | 0.018 | 0.118 |
| VAE | 0.450 | 0.680 | 0.796 | 0.349 | 0.547 | 0.257 | 0.272 |
| GMM | 0.426 | 0.713 | 0.737 | 0.393 | 0.388 | 0.398 | 0.225 |
| DeepSVDD | 0.375 | 0.610 | 0.775 | 0.279 | 0.442 | 0.204 | 0.184 |
| PCA | 0.373 | 0.612 | 0.728 | 0.357 | 0.360 | 0.354 | 0.185 |
| IForest | 0.347 | 0.635 | 0.701 | 0.295 | 0.297 | 0.292 | 0.105 |
| ECOD | 0.340 | 0.637 | 0.720 | 0.345 | 0.345 | 0.345 | 0.167 |
| COPOD | 0.328 | 0.627 | 0.703 | 0.270 | 0.284 | 0.257 | 0.084 |
| SOS | 0.308 | 0.524 | 0.705 | 0.264 | 0.283 | 0.248 | 0.081 |

[7] Note how much more area expensive using STDP synapses that enable online unsupervised learning is compared to much larger (synapse-wise) static synapses in our MNIST accelerator.

### 3.6.1 Discussion: Generalizable Neuromorphic Processing Elements

We have shown how we can use Syn2Logic and its more advanced language constructs to describe hierarchical neuromorphic modules. We showed how the same generic module can be used – through unsupervised learning – for the OPS-SAT benchmark. Our results contribute to and extend the results in OPS-SAT [67] by providing results for unsupervised STDP learning, revealing that the Hebbian-based STDP system generated through the Syn2Logic framework can be a viable (second-best) alternative for inclusion in future CubeSats for on-board unsupervised learning and analysis. Larger systems can easily be generated; in fact, the Syn2Logic description for the explored module can easily be scaled up by changing a few constants, subject only to limitations with the current backend and the size of the available FPGAs.

## 4 Conclusion

In this paper, we have presented the first complete **eNDA** flow, allowing push-button end-to-end generation of neuromorphic systems. Furthermore, we have introduced Syn2Logic— an **eNDA** prototype system implementing the entire **eNDA** flow- and created a class-4 backend targeting the creation of digital neuromorphic systems. We evaluated our proposed **eNDA** flow and prototype implementation on a set of well-known neuron and synapse models and proceeded to adopt our methodology to push state-of-the-art in four fronts: **(i)** Possibly the fastest simulation of the *C. elegans* level-A nematode model for computational neuroscience, **(ii)** a Sudoku accelerator that (for the first time ever) is capable of solving puzzles faster than generic optimization solvers (but does not yet reach tdoku-levels), generalizes to the TOP1465 puzzles set, and is many orders of magnitude faster than prior neuromorphic solvers, and **(iii)** world's most energy-efficient spike-based MNIST accelerator, reaching 5.6 million frames-per-second per Watt spent on a 12-year-old toy-FPGA, with ASIC estimates at >1 billion frames-per-second per Watt, and **(iv)** a generic neuromorphic accelerator capable of online unsupervised learning, reaching second place in terms of $AUC_{PR}$ on the unsupervised space telemetry OPS-SAT benchmark.

**A reflection on generative AI**: An interesting experience is also in the use of generative AI for transporting models between different neural descriptions. One problem in neuroscience is the lack of a common language for describing neural networks (Brian2, NestML, NeuroML, SpineML, Syn2Logic, all use different grammars). We found that using generative AI worked remarkably well to move between different languages, as long as proper validation suites were present (e.g., generating traces using Brian2, moving the model to Syn2Logic using generative AI, and then verifying using traces). So in the future, generative AI could be the very component that unifies cross-framework development between these modeling frameworks.

**Limitations and Future Work**: Future work should cover some of the limitations of the backend. Our backend lays out all neuromorphic components directly *in silico*, which is well within the spirit of digital implementations of traditionally analog neuromorphic systems [57]; our backend – while being extremely performant – cannot easily handle a large number of neurons/synapses before running out of silicon real estate (order $10^3$ neurons on mid-tier FPGAs, possibly reaching $10^4$ neurons on high-end FPGAs)– not entirely unlike their analog counterparts. We see three directions. The first direction could thus be to extend the class4 backend to operate not only in 2D, but also towards 3D monolithic integrated circuits, which could provide more capacity and synergize well with the sparse, biologically plausible nature of spiking neural networks that could assist in the troublesome thermal gradients in such systems. A second direction would be to create a backend that temporally shares the data paths across states, which is in line with most prior digital neuromorphic systems; such a backend would have significantly more capacity, limited only by the amount of external (or on-chip) RAM resources, but would also be slower and less deterministic, and could lose some of the edge that the current backend provides. A third, and maybe most radical, direction is to create a backend that maps the ODEs and model/network structures to a (possibly mixed-signal) analog/memristive technology, allowing exploration of future post-Moore substrates from the **eNDA** flow without changing the descriptions. Finally, we would like to extend the Syn2Logic language to include **(i)** the delaying of signals at the net level, which is a feature often needed but currently lacking, and **(ii)** support for connection blocks (internally and externally controllable), which would allow for external- or internally controlled structural plasticity.

# 5 Methods

## 5.1 Syn2Logic Implementation

The Syn2Logic compiler consists of 17500 lines of C/C++ code that was written manually by the author. The lexical analysis is performed using FLEX, and we created a hand-written recursive-descent parser. Together, these are 2600 lines of program code, and the remaining program code is for the intermediate optimizations and backend generation.

## 5.2 Fixed-Point Representation

All Syn2Logic generated code used a fixed-point representation, represented by two integers $M$ and $N$ in the form Q$M$:$N$, where $M$ are the number of integer bits and $N$ the number of fractional bits, giving a total word length of $M + N + 1$ bits, a representable range of $\pm 2^M$ and a quantisation step of $2^{-N}$.

## 5.3 Neuron Models Exploration

### 5.3.1 Leaky Integrate-and-Fire (LIF)

The LIF neuron integrates its membrane potential $V$ according to the linear subthreshold dynamics

$$C_m \frac{dV}{dt} = g_L (E_L - V) + I, \qquad g_L = \frac{C_m}{\tau_m}, \tag{1}$$

and emits a spike when $V$ crosses a fixed threshold $V_{\text{th}}$, after which $V$ is reset instantaneously to $V_{\text{reset}}$. The parameters are a membrane capacitance $C_m = 250$ pF, a membrane time constant $\tau_m = 10$ ms (so that $g_L = C_m/\tau_m = 25$ nS), a leak and resting potential $E_L = -70$ mV, a threshold $V_{\text{th}} = -55$ mV, and a reset $V_{\text{reset}} = -70$ mV. Brian2 integrates this equation analytically (exact integration) and Syn2Logic uses `LIF` with a membrane resistance $R_{\text{mem}} = \tau_m/C_m = 40\,\text{M}\Omega$, which reproduces the reference leak conductance, retaining the library-default threshold and reset of −55 mV and −70 mV.

### 5.3.2 Izhikevich

The Izhikevich neuron is a dimensionless two-variable model coupling a membrane-like variable $v$ to a recovery variable $u$,

$$\frac{dv}{dt} = 0.04\, v^2 + 5\, v + 140 - u + I, \tag{2}$$

$$\frac{du}{dt} = a\,(b\, v - u), \tag{3}$$

with the spike-and-reset rule

$$\text{if } v \geq 30\,\text{mV}: \quad v \leftarrow c, \quad u \leftarrow u + d. \tag{4}$$

Here the injected current $I$ is the model's native dimensionless drive rather than a physical current in picoamperes. We use the canonical regular-spiking parameters $a = 0.02$, $b = 0.2$, $c = -65$ mV, and $d = 8$, with a peak/cut-off of 30 mV. Brian2 integrates with forward Euler and initialises $v(0) = c$, $u(0) = b\,c$ and Syn2Logic uses `Izhikevich` with the matching library-default coefficients.

### 5.3.3 Adaptive exponential integrate-and-fire (AdEx)

The AdEx neuron augments an exponential spike-initiation term with a slow adaptation current $w$,

$$C_m \frac{dV}{dt} = g_L (E_L - V) + g_L \Delta_T \exp\left(\frac{V - V_T}{\Delta_T}\right) - w + I, \tag{5}$$

$$\tau_w \frac{dw}{dt} = a\,(V - E_L) - w, \tag{6}$$

with the reset, applied when $V$ exceeds the cut-off $V_{\text{peak}}$,

$$\text{if } V > V_{\text{peak}}: \quad V \leftarrow V_{\text{reset}}, \quad w \leftarrow w + b. \tag{7}$$

The parameters are $C_m = 281$ pF, $g_L = 30$ nS, $E_L = -70.6$ mV, an exponential onset threshold $V_T = -50.4$ mV, a slope factor $\Delta_T = 2$ mV, a subthreshold adaptation conductance $a = 4$ nS, an adaptation time constant $\tau_w = 144$ ms, a spike-triggered adaptation increment $b = 80.5$ pA, a reset $V_{\text{reset}} = -60$ mV, and a spike cut-off $V_{\text{peak}} = 0$ mV. Brian2 integrates with forward Euler from $w(0) = 0$ and Syn2Logic uses `AdEx` with the matching library defaults. The Syn2Logic model emits a spike when $V$ crosses $V_T$; because the exponential term diverges within a single step once $V > V_T$, this is numerically equivalent to the $V_{\text{peak}} = 0$ cut-off used by the references.

### 5.3.4 Hodgkin-Huxley Model

The classic conductance-based HH neuron resolves the action potential explicitly through voltage-gated sodium and potassium channels,

$$C_m \frac{dV}{dt} = I - g_{\text{Na}}\, m^3 h\,(V - E_{\text{Na}}) - g_{\text{K}}\, n^4\,(V - E_{\text{K}}) - g_L\,(V - E_L), \tag{8}$$

where each gating variable $x \in \{m, h, n\}$ relaxes according to

$$\frac{dx}{dt} = \alpha_x(V)\,(1 - x) - \beta_x(V)\,x, \tag{9}$$

with the standard textbook rate functions (voltage $V$ in millivolts, rates in $\text{ms}^{-1}$, resting potential $\approx -65\,\text{mV}$)

$$\alpha_m(V) = \frac{0.1\,(V + 40)}{1 - e^{-(V+40)/10}}, \qquad \beta_m(V) = 4\,e^{-(V+65)/18}, \tag{10}$$

$$\alpha_h(V) = 0.07\,e^{-(V+65)/20}, \qquad \beta_h(V) = \frac{1}{1 + e^{-(V+35)/10}}, \tag{11}$$

$$\alpha_n(V) = \frac{0.01\,(V + 55)}{1 - e^{-(V+55)/10}}, \qquad \beta_n(V) = 0.125\,e^{-(V+65)/80}. \tag{12}$$

The parameters are $C_m = 100\,\text{pF}$, a maximum sodium conductance $g_{\text{Na}} = 12\,000\,\text{nS}$ ($12\,\mu\text{S}$), a maximum potassium conductance $g_{\text{K}} = 3600\,\text{nS}$ ($3.6\,\mu\text{S}$), a leak conductance $g_L = 30\,\text{nS}$, and reversal potentials $E_{\text{Na}} = 50\,\text{mV}$, $E_{\text{K}} = -77\,\text{mV}$, and $E_L = -54.402\,\text{mV}$. Brian2 integrates with the exponential-Euler method from the resting steady state ($m = 0.0529$, $n = 0.3177$, $h = 0.5961$) and Syn2Logic uses `HH` with the matching library defaults, detecting a spike on the upward zero-crossing of $V$ via a re-arming threshold flag that guarantees one count per action potential.

### 5.3.5 Numerical Integration

The LIF, Izhikevich, and AdEx models are integrated at $\Delta t = 0.1$ ms on all Brian2 and Syn2Logic platforms. The HH model uses 0.02 ms for both frameworks.

### 5.3.6 Syn2Logic configuration

For the initial correctness evaluation, Syn2Logic was invoked to generate hardware with a fixed-point representation of Q32:24. Fused operations were enabled, and constant rounding was disabled.

### 5.3.7 Stimulation Protocol

The injected current is swept over 0 to 600 pA for the LIF model, 0 to 15 (dimensionless) for the Izhikevich model, 0 to 900 pA for the AdEx model, and 0 to 1000 pA for the HH model. The software references sample these ranges densely (a current step of 5 pA, or 0.1 for the dimensionless Izhikevich drive), whereas the Syn2Logic sweeps use a coarser step (20 pA, or 0.5 for Izhikevich) because each Syn2Logic point is a full compile-and-simulate cycle (using QuestaSim).

### 5.3.8 Precision sensitivity analysis.

Fidelity between implementations quantified with the root-mean-square error between the two curves (test and reference), normalized by the dynamic range (swing) of the reference to make the metric dimensionless and comparable across models with widely different firing-rate scales:

$$\text{NRMSE} \;=\; 100 \times \frac{\sqrt{\dfrac{1}{N}\displaystyle\sum_{i=1}^{N}\left(f_i^{\text{S2L}} - f_i^{\text{ref}}\right)^2}}{\max_i f_i^{\text{ref}} - \min_i f_i^{\text{ref}}} \quad [\%], \tag{13}$$

where $f_i^{\text{ref}}$ and $f_i^{\text{S2L}}$ are the reference and Syn2Logic firing rates at the $i$-th of $N$ current points (points at which either curve is undefined are excluded). NRMSE is reported as a percentage of the reference swing, and a configuration is deemed *faithful* when it meets a 5% design budget.

## 5.4 Synapse Models

### 5.4.1 Plasticity models.

Four canonical learning rules spanning the principal families of synaptic plasticity were studied: **(i)** pair-based spike-timing-dependent plasticity (PairSTDP), **(ii)** triplet STDP, which augments the pair rule with a second-order postsynaptic trace; **(iii)** a reward-modulated ("three-factor") STDP rule (RSTDP) after Izhikevich[68], in which pre/post coincidences accumulate a decaying synaptic eligibility trace that is converted into a weight change only upon delivery of a global reward signal; and **(iv)** the Bayesian Confidence Propagation Neural Network (BCPNN) synapse after Lansner and Tully *et al.*[69] [8]. The synaptic weight was initialised to $w(0) = 0.5$ (dimensionless) in the three STDP-family rules.

### 5.4.2 Pair-based STDP.

Two exponentially decaying eligibility traces, $x_{\text{pre}}$ and $x_{\text{post}}$, record recent pre- and postsynaptic activity,

$$\frac{dx_{\text{pre}}}{dt} = -\frac{x_{\text{pre}}}{\tau_+}, \qquad \frac{dx_{\text{post}}}{dt} = -\frac{x_{\text{post}}}{\tau_-}, \tag{14}$$

with the spike-triggered updates

$$\text{pre-spike:}\ \ x_{\text{pre}} \leftarrow x_{\text{pre}} + 1, \qquad w \leftarrow w - A_-\,x_{\text{post}}, \qquad g_{\text{out}} \leftarrow g_{\text{scale}}\,w, \tag{15}$$

$$\text{post-spike:}\ \ x_{\text{post}} \leftarrow x_{\text{post}} + 1, \qquad w \leftarrow w + A_+\,x_{\text{pre}}. \tag{16}$$

Parameters: $\tau_+ = \tau_- = 20\,\text{ms}$, $A_+ = A_- = 0.01$, output scale $g_{\text{scale}} = 5$.

### 5.4.3 Triplet STDP.

The triplet rule [70] adds slow pre- and postsynaptic traces $r_2, o_2$ to the fast pair traces $r_1, o_1$,

$$\frac{dr_1}{dt} = -\frac{r_1}{\tau_+}, \quad \frac{dr_2}{dt} = -\frac{r_2}{\tau_x}, \quad \frac{do_1}{dt} = -\frac{o_1}{\tau_-}, \quad \frac{do_2}{dt} = -\frac{o_2}{\tau_y}, \tag{17}$$

with the triplet-modulated updates

$$\text{pre-spike:}\ \ w \leftarrow w - o_1\,(A_2^- + A_3^-\,r_2), \qquad r_1 \leftarrow r_1 + 1,\ \ r_2 \leftarrow r_2 + 1, \tag{18}$$

$$\text{post-spike:}\ \ w \leftarrow w + r_1\,(A_2^+ + A_3^+\,o_2), \qquad o_1 \leftarrow o_1 + 1,\ \ o_2 \leftarrow o_2 + 1, \tag{19}$$

and $g_{\text{out}} = g_{\text{scale}}\,w$ emitted on each pre-spike. Parameters: fast decays $\tau_+ = 16.8\,\text{ms}$, $\tau_- = 33.7\,\text{ms}$; slow decays $\tau_x = 101\,\text{ms}$, $\tau_y = 125\,\text{ms}$; pair amplitudes $A_2^+ = 0$, $A_2^- = 0.0072$; triplet amplitudes $A_3^+ = 0.0062$, $A_3^- = 0$; $g_{\text{scale}} = 1$.

### 5.4.4 Reward-modulated STDP.

RSTDP [68] is a three-factor rule: the same pre/post traces feed a slowly decaying eligibility trace $e$,

$$\frac{dx_{\text{pre}}}{dt} = -\frac{x_{\text{pre}}}{\tau_+}, \quad \frac{dx_{\text{post}}}{dt} = -\frac{x_{\text{post}}}{\tau_-}, \quad \frac{de}{dt} = -\frac{e}{\tau_e}, \tag{20}$$

with

$$\text{pre-spike:}\ \ x_{\text{pre}} \leftarrow x_{\text{pre}} + 1, \qquad e \leftarrow e - A_-\,x_{\text{post}}, \qquad g_{\text{out}} \leftarrow g_{\text{scale}}\,w, \tag{21}$$

$$\text{post-spike:}\ \ x_{\text{post}} \leftarrow x_{\text{post}} + 1, \qquad e \leftarrow e + A_+\,x_{\text{pre}}, \tag{22}$$

so that spike timing only builds eligibility; a separate reward event converts it into a weight change,

$$\text{reward:}\quad w \leftarrow w + R\,e, \tag{23}$$

and the reward does not reset $e$. Parameters: $\tau_+ = \tau_- = 20\,\text{ms}$, eligibility decay $\tau_e = 1000\,\text{ms}$, $A_+ = A_- = 0.01$, reward magnitude $R = 1.0$, $g_{\text{scale}} = 1$.

[8] Reference Brian2 code given by Anders Lansner, Pawel Herman, and Ferdinand Bujanowski.

### 5.4.5 BCPNN.

The BCPNN synapse [71] in which low-pass-filtered spike ($Z$), eligibility ($E$) and probability ($P$) traces estimate pre-, post- and joint firing probabilities, and the synaptic weight is the log-odds

$$w = \log \frac{P_{ij}}{P_i P_j}. \tag{24}$$

Each input spike sets a one-timestep pulse $\mathrm{spk}_i$ that decays with the spike duration, $d\,\mathrm{spk}_i/dt = -\mathrm{spk}_i/t_{\mathrm{spike}}$ (and likewise $\mathrm{spk}_j$), driving the spike ($Z$), eligibility ($E$) and probability ($P$) traces

$$\frac{dZ_i}{dt} = \frac{\mathrm{spk}_i/(f_{\max}\,t_{\mathrm{spike}}) - Z_i + \varepsilon}{\tau_z}, \qquad \frac{dZ_j}{dt} = \frac{\mathrm{spk}_j/(f_{\max}\,t_{\mathrm{spike}}) - Z_j + \varepsilon}{\tau_z}, \tag{25}$$

$$\frac{dE_i}{dt} = \frac{Z_i - E_i}{\tau_e}, \quad \frac{dE_j}{dt} = \frac{Z_j - E_j}{\tau_e}, \qquad \frac{dE_{ij}}{dt} = \frac{Z_i Z_j - E_{ij}}{\tau_e}, \tag{26}$$

$$\frac{dP_i}{dt} = \frac{E_i - P_i}{\tau_p}, \quad \frac{dP_j}{dt} = \frac{E_j - P_j}{\tau_p}, \qquad \frac{dP_{ij}}{dt} = \frac{E_{ij} - P_{ij}}{\tau_p}, \tag{27}$$

The drive term $1/(f_{\max}\,t_{\mathrm{spike}}) = 500$ is the largest internal quantity in any of the four rules and sets this synapse's integer-bit floor. Parameters: $f_{\max} = 20\,\mathrm{Hz}$, floor $\varepsilon = 0.05$, $t_{\mathrm{spike}} = 0.1\,\mathrm{ms}$, $\tau_z = 10\,\mathrm{ms}$, $\tau_e = 100\,\mathrm{ms}$, $\tau_p = 1000\,\mathrm{ms}$; marginal traces initialised to $\varepsilon$ and joint traces to $\varepsilon^2$.

### 5.4.6 Induction protocols.

Each rule was driven by a deterministic spike protocol designed to exercise its characteristic dynamics. PairSTDP and TripletSTDP received a mixed pairing protocol combining potentiating and depressing pre/post orderings across repeated bursts. RSTDP received a four-phase protocol (causal pairing with reward; causal pairing without reward; anti-causal pairing with reward; causal pairing followed by a delayed reward), with a quiescent washout longer than twice the eligibility time constant between phases so that each phase began from a null eligibility trace. BCPNN received the five-section correlation protocol of Tully *et al.* (correlated, independent, anti-correlated, silent, and postsynaptically-muted epochs; 10 s total, $\Delta t = 0.1$ ms, peak rate 20 Hz), with spike trains drawn from a seeded pseudo-random generator so that hardware and reference implementations received bit-identical input. Protocols consisting only of binary spike events are independent of the fixed-point format and were generated once; the RSTDP protocol additionally carries an analogue reward payload whose port width scales as $\mathrm{Q}M{:}N$, and was therefore regenerated for every format.

### 5.4.7 Cycle-accurate simulation.

Each compiled model was co-generated with a self-checking VHDL testbench that applies the protocol one row per clock cycle (one clock cycle corresponds to one biological timestep) and records every state variable each cycle. Testbenches were compiled and simulated cycle-accurately in QuestaSim. Timesteps were 0.01 ms.

### 5.4.8 Precision sensitivity analysis.

For every rule we recorded the full weight trajectory $w(t)$ at each fixed-point format and quantified its fidelity relative to a high-precision reference format ($Q_{16:16}$ for the STDP-family; $Q_{24:24}$ for R-STDP and BCPNN). Fidelity was expressed as the normalised root-mean-square error of $w(t)$ against the reference $w_{\mathrm{ref}}(t)$,

$$\mathrm{NRMSE} \;=\; \frac{\sqrt{\dfrac{1}{|\mathcal{T}|}\displaystyle\sum_{t\in\mathcal{T}}\big(w(t) - w_{\mathrm{ref}}(t)\big)^2}}{\max_t w_{\mathrm{ref}}(t) - \min_t w_{\mathrm{ref}}(t)} \times 100\%, \tag{28}$$

normalised by the reference weight swing and reported as a percentage. The set $\mathcal{T}$ comprises only those cycles at which both trajectories are finite, guarding against fixed-point underflow of the BCPNN $P$-traces (for which the log-odds in Eq. (24) are undefined). The $(M, N)$ plane was first mapped at coarse resolution to locate the operating region and subsequently at single-bit resolution over the relevant range to resolve fine structure. From each map we extracted, for a fixed error budget (5% NRMSE), the smallest total word length $M + N + 1$ achieving that budget, and the minimum integer width below which the largest internal trace of each rule saturates.

## 5.5 Caenorhabditis elegans

### 5.5.1 Reference network.

We evaluate our neuromorphic compilation flow on the *C. elegans* connectome, using the `c302` *Level A* model as the ground-truth specification. The network comprises 302 neurons with the full set of connections defined by `c302`. All neurons are integrate-and-fire cells driven by a uniform tonic input of 2.1 pA with a flat conductance, the operating point at which the model produces stable, non-degenerate activity across the whole network. It is transcribed into the Syn2Logic domain-specific language as a single flat network, with a uniform synaptic conductance.

### 5.5.2 Software baselines.

To establish a reference spike train we simulate the same network with three independent simulators: jNeuroML (the LEMS reference interpreter), Brian2, and NEST (32 OpenMP threads). All three integrate for 1000 ms at a fixed time step of 0.1 ms (10k steps).

### 5.5.3 Hardware compilation.

Syn2Logic compiles the `.s2l` description directly to synthesizable VHDL. The generated datapath advances the entire network by one simulation time step every clock cycle, so the 1000 ms experiment maps to exactly 10 000 clock cycles independent of firing activity. Two orthogonal knobs reduce hardware cost while preserving accuracy: **(i)** a fixed-point number format, here Q7:15 (7 integer, 15 fractional bits) with non-fused arithmetic; and **(ii)** a synapse-pruning factor `pmw`, set to 0.2, which only propagates the 20 % active synaptic weights into the neurons. Constant coefficients are additionally quantized with up to 1 % relative error (`--round-constants=0.01`). We verify the emitted RTL bit-exactly against the compiler's fixed-point model in Siemens QuestaSim, using a testbench that counts per-neuron spike events; per-neuron firing rates correlate with Brian2 at r = 0.9967.

## 5.6 Sudoku Accelerator

### 5.6.1 Puzzles used and Performance Measurements

Tuning our SNN accelerator was performed on all 46 puzzles proposed by Mantere–Koljonen [63]. Generalization was tested using the TOP1465 [64] and Euler 96 [72]. We removed the overlapping puzzles between Mantere-Koljonen and Euler 96.

For performance testing, we only counted time to solve the puzzle and not application binary/hardware startup time or puzzle configuration (including the setup time would skew the results significantly in favor of the single-cycle neuromorphic configuration). For CP-SAT, we tested the strong scalability on the puzzles and used the single worker version, which was the fastest.

### 5.6.2 Solvers used

We compared our neuromorphic solvers to three other state-of-the-art methods. SCIP (Solving Constraint Integer Programs) is a widely used open-source solver for mixed-integer programming (MIP) and mixed-integer nonlinear programming (MINLP), developed primarily at the Zuse Institute Berlin. It combines branch-and-bound, cutting planes, and constraint programming in a plugin-based framework that lets users customize components like branching rules, heuristics, and constraint handlers.

CP-SAT is Google's open-source constraint programming solver, distributed as part of the OR-Tools suite. It reformulates integer and combinatorial problems as boolean satisfiability (SAT) problems and solves them using a clause-learning SAT engine combined with linear programming relaxations, constraint propagation, and large neighborhood search, running multiple strategies in parallel.

Tdoku is a highly optimized open-source Sudoku solver written in C++ by Tom Dillon, designed to be among the fastest in existence for both easy and hard puzzles. It models Sudoku as a constraint satisfaction problem using a compact box-band representation and applies vectorized with SIMD instructions (SSE/AVX2/AVX-512) for speed.

### 5.6.3 Sudoku Neuron model

Each of the $9 \times 9 \times 9 = 729$ neurons $B_{x,y,z}$ (cell $(x, y)$ asserting digit $z{+}1$) is an Izhikevich unit augmented with two per-neuron, exponentially decaying *escape currents* that let the winner-take-all (WTA) network leave local minima. $v$ is the membrane variable and $u$ for the recovery variable, a *free* neuron (clue = 0) updates as (please note how we compute $(0.04 * v) * v$ and not $0.04 * v^2$ to prevent product blow-up):

$$\dot{v} = (0.04 * v) * v + 5v + 140 - u + \big(I_{\text{const}} - I_{\text{syn}} - \theta - r\big), \tag{29}$$

$$\dot{u} = a\,(b\,v - u), \tag{30}$$

$$\dot{\theta} = -\theta/\tau_\theta, \qquad \dot{r} = -r/\tau_r, \qquad \dot{I}_{\text{syn}} = -I_{\text{syn}}/\tau_s. \tag{31}$$

$\theta$ is the reverse adaptation current, and $r$ is a slower reverse current– both use as an annealer within the neuron. Inhibition $I_{\text{syn}}$ enforces the four Sudoku constraints: whenever a conflicting neuron fires, $I_{\text{syn}} \leftarrow I_{\text{syn}} + w_{\text{syn}}\,n_c$, where $n_c$ (described in Syn2Logic using the `::num` reducer) is the number of *simultaneously* conflicting spikes (row, column, box, and one-digit-per-cell). A spike is emitted when $v \geq v_{\text{th}} = 32$, upon which

$$v \leftarrow c = -65, \qquad u \leftarrow u + d, \qquad \theta \leftarrow \theta + \Delta\theta, \tag{32}$$

followed by the floor $v \leftarrow \max(v, c)$ that prevents the $0.04v^2$ positive-feedback well from re-diverging under strong inhibition. A *clue* (given) neuron is pinned: it is driven by $\dot{v} = 0.04v^2 + 5v + 140 - u + I_{\text{set}}$ and is immune to $I_{\text{syn}}$, $\theta$, and $r$. The system is integrated by forward Euler with step $\Delta t$ (also a tunable parameter).

### 5.6.4 Parameter search

Finding parameters was key to a good Sudoku solver, and consumed nearly a week of overnight (iterative) searches. The dynamics are governed by several tunable parameters, $\Theta = \{a, b, d, \Delta t,\ I_{\text{const}}, I_{\text{set}}, w_{\text{syn}}, \tau_s,\ \Delta\theta, \tau_\theta, R_{\text{amp}}, \tau_r\}$. We optimize the single fixed set that solves the largest number of the 46 Mantere–Koljonen benchmark puzzles [63] under a per-puzzle step cap, breaking ties by minimum total step count. Two properties of the landscape dictate the method. First, because $v$ evolves in single-precision floating-point through a quadratic map, the good operating points are *razor-thin, isolated spikes* (floating-point chaos), not smooth basins: gradient- or edge-pushing searches, and hill-climbing from a known-good point, repeatedly fail—nudging a single parameter off a winning value collapses the score. Second, evaluation is cheap and embarrassingly parallel. We therefore search by randomised (Latin-hypercube-style) sampling over all twelve parameters at once, rather than refining locally, and validate any directional hypothesis with a cheap single-axis anchor probe before committing a sweep. The harness compiles the reference solver into an environment-tunable binary, evaluates all 46 puzzles per configuration, and tags every result line with its full parameter vector; runs are resumable and use fail-fast pruning (a configuration is abandoned as soon as it can no longer beat the incumbent). Over $5.5 * 10^4$ configurations, this yields a single set solving all 46 with no per-puzzle retuning. Deploying on hardware is a second stage: the fixed-point datapath (Q8:2) is a *different* dynamical system, so the operating point is *re-optimised* in the same way at the target precision rather than merely quantized from the floating-point solution.

## 5.7 S-MLP Accelerator

### 5.7.1 Training

Our MNIST accelerator implements a fully connected spiking multilayer perceptron 784 – 128 – 64 – 10 (109 184 parameters, no biases). Neurons are non-leaky integrate-and-fire ($\beta = 1$, threshold 1, reset-by-subtraction). Inputs are rate-coded: each pixel intensity is the Bernoulli firing probability of its input neuron, redrawn every timestep, and the class decision is the output neuron with the most spikes over $T$ (25 or 50) timesteps.

The network was trained using snnTorch [36] with backpropagation through time and a fast-sigmoid surrogate gradient on the standard MNIST split [73]. Hardware constraints are applied during training rather than after it: the network is first pruned by weight magnitude and fine-tuned for five epochs at a fixed mask, and the per-neuron bandwidth limit is then fine-tuned for a further four epochs *with the hardware selection in the loop*, so the trained weights are the ones the generated logic actually uses. All accuracies are the full 10 000-image test set; repeated evaluation of one checkpoint can vary by ±0.3 points with the Poisson seed.

### 5.7.2 Power Measurement and Estimation

FPGA power is measured on the DE10-Lite (MAX10) board using a Siglent SDM4055A multimeter for the full System-on-Chip, which includes (except the Syn2Logic-generated accelerator) a JTAG interface, a JTAG command FSM, a Phase-Locked Loop (PLL), and a spike buffer holding one frame worth of data. A 0.01 Ω sense resistor (`R22`) in the 1.2 V core rail of the MAX10 device drops 2.561 mV while the accelerator classifies, giving 256.1 mA and 0.307 W. This is the whole system-on-chip on that rail, including the static leakage of the entire device, and is therefore an upper bound on the classifier alone. For ASIC, we estimated the power using Cadence Voltus for the entire place-and-routed netlist at 0.7 V with a 500 MHz clock.

### 5.7.3 State-of-the-Art

We compared against the following works: DeepFire2 [48], Sommer et al. [74], SyncNN [75], E3NE [76], SeaSNN [77], Fang et al. [78], FPGA-NHAP [79], Ju et al. [80], Spiker+ [50],Firefly v2 [81], and Minitaur [82] on FPGA, and against TrueNorth [83], Loihi [21], and SpiNNaker [84] on ASIC. We used performance numbers taken from the paper itself and derived metrics only where possible (e.g., throughput and energy reported separately). The results spans work achieving between 92 % and 99.42 % MNIST accuracy on networks sized 17k to 647k parameters.

## 5.8 Satellite Telemetry Anomaly Detection

### 5.8.1 Task, encoding and readout

OPS-SAT-AD [67] provides 2,123 telemetry segments described by 18 statistical features (duration, mean, variance, peak counts, smoothed and differenced statistics, . . . ), split by the authors into 1,594 training and 529 test segments (anomaly rate ≈21%). We z-score and min–max normalise every feature to $[0, 1]$ on the training split and present a segment as an 18-channel Bernoulli spike train: channel $q$ fires in each 1 ms step with probability $0.5\, x_q$ for 60 steps, followed by 40 silent steps, so every segment occupies exactly 100 clock cycles. The readout is the protocol the benchmark applies to all its baselines: each neuron is assigned to the *nominal* or *anomaly* class by its row-normalised mean response to the labelled training segments, a segment's anomaly score is the fraction of its spikes emitted by anomaly-assigned neurons, and the binary decision threshold is the $80^{\text{th}}$ percentile of the training scores (PyOD convention, contamination 0.2). Labels are used only for this readout and threshold, never for learning.

### 5.8.2 Neuron model

The core is four columns of 14 positive-only leaky integrate-and-fire (LIF) neurons. All state is dimensionless and non-negative, which makes a 7-bit Q1:5 fixed-point datapath (one sign, one integer and five fractional bits) sufficient. With one Euler step per clock cycle ($\Delta t = 1$ ms of model time), a neuron obeys

$$\tau_m \dot{V} = -V + I_{\text{syn}} - I_{\text{inh}}, \qquad \tau_s \dot{I}_{\text{syn}} = -I_{\text{syn}}, \qquad \tau_s \dot{I}_{\text{inh}} = -I_{\text{inh}}, \tag{33}$$

with $\tau_m = 10$ ms and $\tau_s = 2$ ms. Every presynaptic event adds the synapse's current to $I_{\text{syn}}$. Lateral inhibition is *counted*: when $n$ neurons of the same column fire in a cycle, $I_{\text{inh}} \leftarrow I_{\text{inh}} + C_{\text{inh}}\, n$ ($C_{\text{inh}} = 1.6$); the merged spike of each other columns adds the weaker $C_{\text{ext}}\, n$ ($C_{\text{ext}} = 0.16$), which decorrelates the prototype sets of the four columns. A spike is emitted when $V > 1$ and resets $V \leftarrow 0$. Because $I_{\text{inh}}$ saturates at the Q1:5 rail (1.97), the inhibition acts as a hard winner-take-all; a homeostatic threshold present in the model is provably inert at this precision (its decay underflows) and can be removed from the hardware without changing a single spike.

### 5.8.3 Synapse model and on-chip learning

Each neuron owns 18 plastic synapses, one per feature, of which a fixed connectivity mask enables exactly three (balanced so that every feature reaches 9–10 neurons). Enabled weights start uniformly random in $[0, 0.3]$; the mask is the only symmetry breaker. A masked synapse emits no event and no current, so the 840 masked synapses of the 1,008 could be pruned without effect. The synapse implements soft-bounded pair STDP with presynaptic and postsynaptic traces $x_{\text{pre}}, x_{\text{post}}$ ($\tau_\pm = 6$ ms):

$$\text{pre spike:} \quad x_{\text{pre}} \leftarrow x_{\text{pre}} + 1,\; w \leftarrow w - A_-\, x_{\text{post}}\, w\, \ell,\; I_{\text{out}} = I_{\text{scale}}\, w, \tag{34}$$

$$\text{post spike:} \quad x_{\text{post}} \leftarrow x_{\text{post}} + 1,\; w \leftarrow w + A_+\, x_{\text{pre}}\, (w_{\max} - w)\, \ell, \tag{35}$$

with $A_+ = 0.04$, $A_- = 0.016$, $w_{\max} = 1$, $I_{\text{scale}} = 5$. The gate $\ell \in [0, 1]$ is set to 1 by a learn signal carried with the stimulus and decays with $\tau_\ell = 2$ ms, so plasticity is on during the training pass and off during evaluation without any control logic in the core. Training is unsupervised and takes place entirely on the chip: the 1,594 training segments are streamed once (159,400 cycles) with the learn bit set, after which the same segments and the 529 test segments are streamed with learning off, and the output spike trains are captured.

### 5.8.4 Hardware

The chosen configuration is wrapped in a system-on-chip on a DE10-Agilex board with on-chip stimulus, configuration and capture memories, a clock-gated sequencer and a JTAG host link; one training-plus-evaluation run takes 371,700 cycles, 25 ms at the 15 MHz core clock. The board reproduces the RTL simulation spike-for-spike (56,960 of 56,960 eval spikes). We compare against the 23 unsupervised detectors in the benchmark's Table 3 (PyOD implementations, identical split, protocol, and metrics).

## 5.9 Software

### 5.9.1 NEST

NEST (Neural Simulation Tool) is an open-source simulator for spiking neural network models, designed to study the dynamics, size, and structure of neural systems rather than the detailed morphology of individual neurons. It represents neurons as point or few-compartment models and focuses on the network level, making it well-suited for large-scale simulations involving thousands to millions of neurons and their connections. NEST is optimized for performance and scales efficiently from laptops to high-performance supercomputers through parallel computing (using both threading and MPI). Models are typically built and controlled through PyNEST, its Python interface, while the computationally intensive simulation runs in a fast C++ core. It supports a wide range of neuron and synapse models, including plasticity mechanisms like STDP, and is widely used in computational neuroscience research. In our experiments, we used NEST 3.10.

### 5.9.2 Brian2

Brian2 is an open-source simulator for spiking neural networks, written in Python and designed around flexibility and ease of model specification. Neuron and synapse models are defined directly as systems of differential equations using standard mathematical notation and physical units, allowing custom dynamics, threshold and reset conditions, refractoriness, and synaptic plasticity rules to be expressed without low-level programming. To maintain performance despite this high-level interface, Brian2 uses runtime code generation: model descriptions written in Python are translated into efficient compiled C++ code that is executed at simulation time. The simulator provides a range of numerical integration methods, from exact integration for linear systems to numerical schemes for arbitrary equations, and supports monitoring of spikes, state variables, and population activity. We used Brian2 v2.10.1 in all our experiments.

### 5.9.3 Cadence EDA toolchain

The digital implementation flow was carried out using the Cadence toolchain. Logic synthesis was performed with Cadence Genus v25.10, which takes the RTL description together with timing constraints (SDC) and a standard-cell library and produces an optimized gate-level netlist, performing logic mapping and optimization for timing, area, and power. The resulting netlist was then passed to Cadence Innovus v25.10 for physical implementation (place-and-route), where it is combined with the standard-cell library and physical (LEF) data to generate the final routed layout. The Innovus flow encompassed floorplanning, power planning, placement (via the GigaPlace engine), clock tree synthesis (CCOpt), signal routing (NanoRoute), and concurrent timing and power optimization, yielding a placed-and-routed design from which area, timing, and power metrics were extracted.

### 5.9.4 Siemens Questasim

QuestaSim is a commercial hardware simulator from Siemens EDA (formerly Mentor Graphics) used to verify digital designs written in VHDL, Verilog, or SystemVerilog before they reach silicon or an FPGA.

### 5.9.5 snnTorch

The S-MLP architecture was implemented and trained using snnTorch v0.9.4, which is an open-source Python library that extends the PyTorch framework to spiking neuron models. snnTorch exposes spiking neurons, input encoders, surrogate gradients, and loss functions as native PyTorch modules, allowing SNNs to be constructed and trained using standard deep learning practices and inheriting PyTorch's automatic differentiation and GPU acceleration. Networks are defined by composing spiking neuron layers (such as leaky integrate-and-fire units) with conventional PyTorch layers, and unrolled over discrete time steps to process temporally encoded inputs.

### 5.9.6 Altera Quartus

Altera Quartus is a software toolchain used to design digital circuits for Altera FPGAs and CPLDs. Quartus lets a developer compile their designs, perform timing analysis, examine RTL diagrams, simulate the design's reaction to different stimuli, and configure the target device with the programmer. We used several versions: Quartus Pro for the Terasic DE10-Agilex platform, while Quartus Lite 18.1 was used for the DE10-Lite MAX10 board.

## 5.10 Experimental Platform

All experiments were carried out on a server featuring two Intel Xeon Silver 4514Y processors, a total of 32 cores (64 hyperthreads) and 256 GB of DDR5 RAM. The system is running Rocky 9.8 with kernel 5.14.0-687.10.1. All performance-sensitive experiments on the server were executed in isolation using Slurm 25.11.4. For FPGA execution, we used a mid-range Terasic DE10-Agilex featuring an Agilex 7 (AGFB014R24) (1.4M LEs, 139 Mb SRAM, 4510 DSPs), as well as a low-range Terasic DE10-Lite featuring a MAX10 10M50 device (50k LEs, 1.6 Mb SRAM, 64 equivalent DSP blocks).

# Supplementary information

**Code Availability.** A Docker image with the compiler, sources, and bitstreams is available to reviewers via the supplementary material.

**Data availability.** All data, scripts, and results have been attached to this paper as supplementary material.

**Competing interests.** The author declares no competing interests.

**Acknowledgements.** This work was supported by the Swedish Research Council's Project Building Digital Brains under Grant 2021-04579. The author would also like to thank Muhammad Ihsan Al Hafiz, Wiktor Jan Szczerek, Björn Lindqvist, Pedro Antunes, Pawel Herman, and Anders Lansner for valuable and interesting discussions on the neuromorphic topic over the years.

# Declarations

The **Syn2Logic prototype framework has been – from start to finish – written by hand by the author**, as has the description of all neural models. The author only used Claude Fable 5.0 and Opus 5.0 for developing Syn2Logic descriptions (`.s2l` files) of networks (some converted from NEST/Brian2 to Syn2Logic), the BCPNN synapse (from Brian2), plotting, scripting, and help in analyzing the experiments. Grammarly/Claude Fable 5.0 was used to polish/improve the quality of the written manuscript.